ASTON BUSINESS SCHOOL

Post Graduate MBA Programme

CANDIDATE NUMBER: 959314

Student Name: James Bell

Title:

# The global economic impact of AI technologies in the fight against financial crime.

Master of Business Administration

SUBMISSION DATE: 19/09/2022

*I declare that I have personally prepared this project and that it has not in whole or in part been submitted for any other degree or qualification. Nor has it appeared in whole or in part in any textbook, journal or any other document previously published or produced for any purpose. The work described here is my/our own, carried out personally unless otherwise stated. All sources of information, including quotations, are acknowledged by means of reference, both in the final reference section, and at the point where they occur in the text.*

8719 Words

**Acknowledgements**

The author would like to acknowledge the support of his supervisor, family, friends, and peers.

## Contents

## Acronyms

| Acronym | Meaning | Notes |
|---|---|---|
| AI | Artificial Intelligence | Umbrella term for the use of computers to simulate activities usually reserved for human-like intelligence. Includes Machine Learning as a technology. |
| AML | Anti-Money Laundering | |
| API | Application programming interface | |
| BERT | Bidirectional Encoder Representations from Transformers | A Google technology for developing language-capable AIs |
| CCO | Chief Compliance Officer | |
| CD | Creative destruction | |
| DMLRO | Deputy Money Laundering Reporting Officer | |
| FATF | Financial Action Task Force | |
| FCA | Financial Conduct Authority | A UK Financial Regulator |
| GDP | Gross Domestic Product | |
| GDPR | General Data Protection Regulation | |
| KYC | Know Your Customer | |
| ML | Money Laundering | |
| MLRO | Money Laundering Reporting Officer | |
| NCA | UK National Crime Agency | |
| OFAC SDN | Office of Foreign Assets Controls "Specially Designated Nationals And Blocked Persons List" | |
| PATRIOT | Uniting and Strengthening America by Providing Appropriate Tools Required to Intercept and Obstruct Terrorism Act | |
| PEP | Politically Exposed Person. | An individual with a prominent public function. |
| POCA | Proceeds of Crime Act of 2002 | |
| SAR | Suspicious Activity Report | |
| SEC | U.S. Securities and Exchange Commission | A US federal agency |
| TRLs | Technology readiness levels | A technology maturity model |
| UBO | Ultimate beneficial Owner | |
| UKFIU | UK Financial Intelligence Unit | |
| UKHMT | The UK Treasury | |
| UNODC | United Nations Office on Drugs and Crime | |

## Executive summary

*Money laundering* is a crime that enables the globalisation of all others. It is fought by teams embedded in companies, and financial organisations worldwide, based on recommendations from a single global watchdog interpreted into local legislation. It is watched over by regulators empowered by laws to punish and fine anyone not correctly performing the proscribed duties and obligations of Anti-money laundering ('AML'). Those who labour to meet the regulations relentlessly screen information coming in and out of their businesses, looking for matches with lists of names or patterns of behaviour within data streams, any indication that their customer might be suspicious. Such suspicions are to be reported to the government immediately. So many are these reports that governmental bodies are deluged in them, unable to sort the self-protective reflex report from the definitive alert for criminal activity. Some companies have thousands of employees performing match reviews daily on millions of data records per hour. The volume of matches to be reported grows ever upwards.

This industry considers Artificial Intelligence ('AI') the possible saviour to the increasing burden. The companies hope to be able to filter out false positives, and the regulators want to be able to automate the analysis of vital criminal intelligence.

Is the rapid adoption of Artificial Intelligence a sign that "creative destruction" (a capitalist innovation process first theorised in 1942) is occurring? Although its theory suggests that it is only visible over time in aggregate, this paper devises three hypotheses to test its presence on a macro level and research methods to produce the required data. This paper tests the theory using news archives, questionnaires, and interviews with industry professionals. It considers the risks of adopting Artificial Intelligence, its current performance in the market and its general applicability to the role.

The results suggest that creative destruction is occurring in the AML industry despite the activities of the regulators acting as natural blockers to innovation. This is a pressurised situation where current-generation Artificial Intelligence may offer more harm than benefit. For managers, this paper's results suggest that safely pursuing AI in AML requires having realistic expectations of Artificial Intelligence's benefits combined with using a framework for AI Ethics.

## Chapter 1. Introduction

### Project aim

This applied research project aims to critically assess the benefits and drawbacks of Artificial Intelligence ('AI') technologies in the fight against money laundering.

### Context

UNODC issued a report in 2011, which estimated that the global proceeds of crime in 2009 accounted for between 2.3 and 5.5 per cent of global GDP ($2.1 trillion) (UNODC, 2011). However, thanks partly to the obscuring effects enabled by money laundering, less than 1% of this staggering amount is seized.

### What is money laundering?

Money laundering ('ML') is a process that disguises the illegal sources of funds. These funds represent a cost in lives brought about by drugs, human trafficking, terrorist financing and arms smuggling. Once illegal funds are "placed" into the financial system, the money is moved around and "layered" through multiple accounts to disguise its source, eventually producing "integrated" funds that appear legal and that can be drawn upon to finance further business. That these illegal and untaxed gains flow around the global financial system has direct consequences on the financial stability of many countries (Aluko and Bagheri, 2012). In addition, ML reduces developing states' ability to meet their citizens' needs. Finally, within the private sector, ML distorts the markets and impacts the reputations of their work sectors (Fundanga, 2003).

Šikman and Grujić note that money laundering is a derivative form of crime, which is to say it is a crime that necessarily results from the outcomes of other criminal activity (Šikman and Grujić, 2021). As such, it can impact a country's economic stability by concealing the source of criminal earnings and financing future crimes (International Monetary Fund, 2016). This was noted by the Financial Conduct Authority ('FCA') (the UK regulator) to cause "incalculable" damage to society and something to which technology might be the answer (Butler, 2019). However, while societal damage remained "incalculable", the FCA did calculate the costs in terms of the number of offences (3.8 million fraud offences in the UK in 2019) and the costs of fighting it (£37bn per year), which gives some level of insight into the fundamental problem. Money laundering can therefore be considered a direct obstacle to the growth of many nations.

**What is the context of anti-money laundering?**

The various country definitions of anti-money laundering were unified by the creation, in 1989, of a G7 watchdog called the Financial Action Task Force ('FATF'), which issues recommendations to combat money laundering.

**Anti-money laundering regime example: UK**

Each region implements the FATF recommendations in their own way. In the UK, this is through Acts of Parliament. An important example is the Proceeds of Crime Act of 2002 ('POCA'), which significantly impacted those fighting money laundering. For example, part 7 of the Act created three types of offences connected with money laundering: the concept of "criminal property", the "failure to report" suspicions of ML, and "tipping off" (telling a third party that they are being investigated). The act makes handling the proceeds of crime itself illegal, which places the onus very much onto the banks to get their procedures right and creates a regulatory "burden" (Legislation.gov.uk, 2002). The money laundering regulations cover how those procedures are defined.

**The MLRO, a focus of pressure on the regulated**

The UK Money Laundering Regulations (2007), recently replaced by The Money Laundering, Terrorist Financing and Transfer of Funds (Information on the Payer) Regulations (2017), defined the Money Laundering Reporting Officer ('MLRO') role within regulated companies. This role oversees controls and procedures, monitoring for criminals, and reporting suspicions to the National Crime Agency ('NCA') by creating a Suspicious Activity Report ('SAR'). The MLRO raising a SAR provides valuable intelligence to the NCA (nationalcrimeagency.gov.uk, 2022) for use in their fight against criminal activity.

A "suspicion" can be made up of several more minor indicators (such as a pattern of behaviour) or one big flag (such as a match on a name found in the sanctions list). For the MLRO, however, raising a SAR is an obligation under which they must freeze the account transaction or potentially face criminal charges themselves. The weight of this pressure on the regulated is perfectly illustrated by the revised guidance in 2021 on prosecuting money laundering offences from the Crown Prosecution Service ('CPS'). Section 330 of the POCA criminalises the "failure to report" money laundering. The new guidance enables the charge against someone for failure to report, even where there is insufficient evidence that any money laundering had been planned. In other words, any unreported suspicion of money laundering is criminal <u>regardless of whether money laundering actually took place.</u>

**Research issues identified by gaps in literature**

The impact of this pressure on the regulated companies, the MLRO and other compliance workers is currently an unexamined gap in the literature:

1. This paper contends that pressure leads to the rapid adoption of Artificial Intelligence technologies for anti-money laundering ('AML') without carefully considering the risks involved.
2. Moreover, this paper will also investigate how the high regulatory burden acts as a restriction or blocker to innovation in the AML market.
3. It may be too high risk to combat money laundering with AI, and a route must be discovered that balances AML effectiveness and the needs of the working populations.

It will establish these arguments by completing the following objectives:

**Objectives: Guide rails for project activities**

1. To place Anti Money Laundering ('AML') into the context of the technological and regulatory approaches used to combat financial crimes (Substantive objective).
2. To critically assess the current regulatory and technological paradigm (Substantive objective).
3. To examine the use of AI technologies in AML and assess their benefits and drawbacks through economic theory (Theoretical objective).
4. To develop a series of hypotheses regarding AI in AML, which can be tested against the data collected (Methodological objective).

This paper will thereby contribute new direct research on the people working in the industry. As such, this paper will be valuable for other researchers looking at these issues and provide a jumping-off point to conduct their research. For practitioners working in AI, this paper will provide insight into the best and safest practices in AI development for AML. Finally, for policymakers, the conclusive chapter will offer policy recommendations for strategically using AI.

All sections of this paper will directly address one or more of the objectives and cohesively develop a set of logical assumptions (the warrant) that connects the evidence collected (the grounds) with the claims made (the claims).

## Chapter 2. Extended literature review

### Introduction

This literature review considers the money laundering issue through the lens of risk. Research by Korystin (O. Y. Korystin *et al.*, 2021) suggests that the further one travels from the financial centres to states with lower technological maturity, the higher the risk is present to economic development from the "shadow economy." Such is the relationship discovered that the higher the level of vulnerability, the higher the spread of money laundering facilitating it. The determination that this vulnerability is related to the effectiveness of AML is outlined by Idowu and Obasan (Idowu and Obasan, 2012). They find a correlation between a bank's performance, economic benefit, and ability to combat money laundering.

AML is a technological arms race between criminals and innovators, and this would follow the observations of Romer (Romer, 2016) that endogenous technological change is a model for positive growth when incentivised. Once created, the technological "instructions" can be replicated without further costs. Only developing a new set of "instructions", called a new “technological paradigm”, incurs the equivalent of a fixed cost (Romer, 2016, p. S72).

Therefore, banks invest heavily in anti-money laundering technologies to alert against suspicious transactions. However, as Bertrand (Maxwell, Bertrand and Vamparys, 2020) noted, these detection systems can cause other problems and produce prodigious volumes of potential alerts, the vast majority of which are false positives but must still be reviewed by a human. For example, in the UK alone, over 460,000 suspicious activity reports ('SARs') are sent to the UK Financial Intelligence Unit ('UKFIU') per year (NCA, 2022), far more than can be productively investigated.

Moreover, any customer transaction must be processed promptly or face the customer's ire. This places the bank in a bind between the regulations (the fines for regulation violations being US$8.14bn globally in 2019 (Burns, 2020)) and being sued by a customer who has potentially only a similar name to a sanctioned person. Unsurprisingly, as noted by Jayasekara (Jayasekara, 2020), many banks close accounts of those mired in high-risk countries to reduce the bank's risk profile. Banks employ large numbers of staff to review potential matches in good time, with Citi Bank counting 30,000 employees in "risk, regulatory

and compliance" roles in 2018 (Citigroup Inc., 2018, p. 54). This number is out of a total employee population of 204,000 and amounts to 14% of global staff performing this duty.

High volumes lead to defensive reporting of suspicions to governmental bodies to head off any potential downside. Described as the "Crying Wolf Theory" by Raweh (Raweh, Erbao and Shihadeh, 2017), this excessive reporting deluges investigative units, who, as noted by Bertrand (Maxwell, Bertrand and Vamparys, 2020), are only able to review up to 20% of the reports they receive (and fewer sent onto law enforcement).

*Figure 1 - History of AML screening technologies. Source: the author*

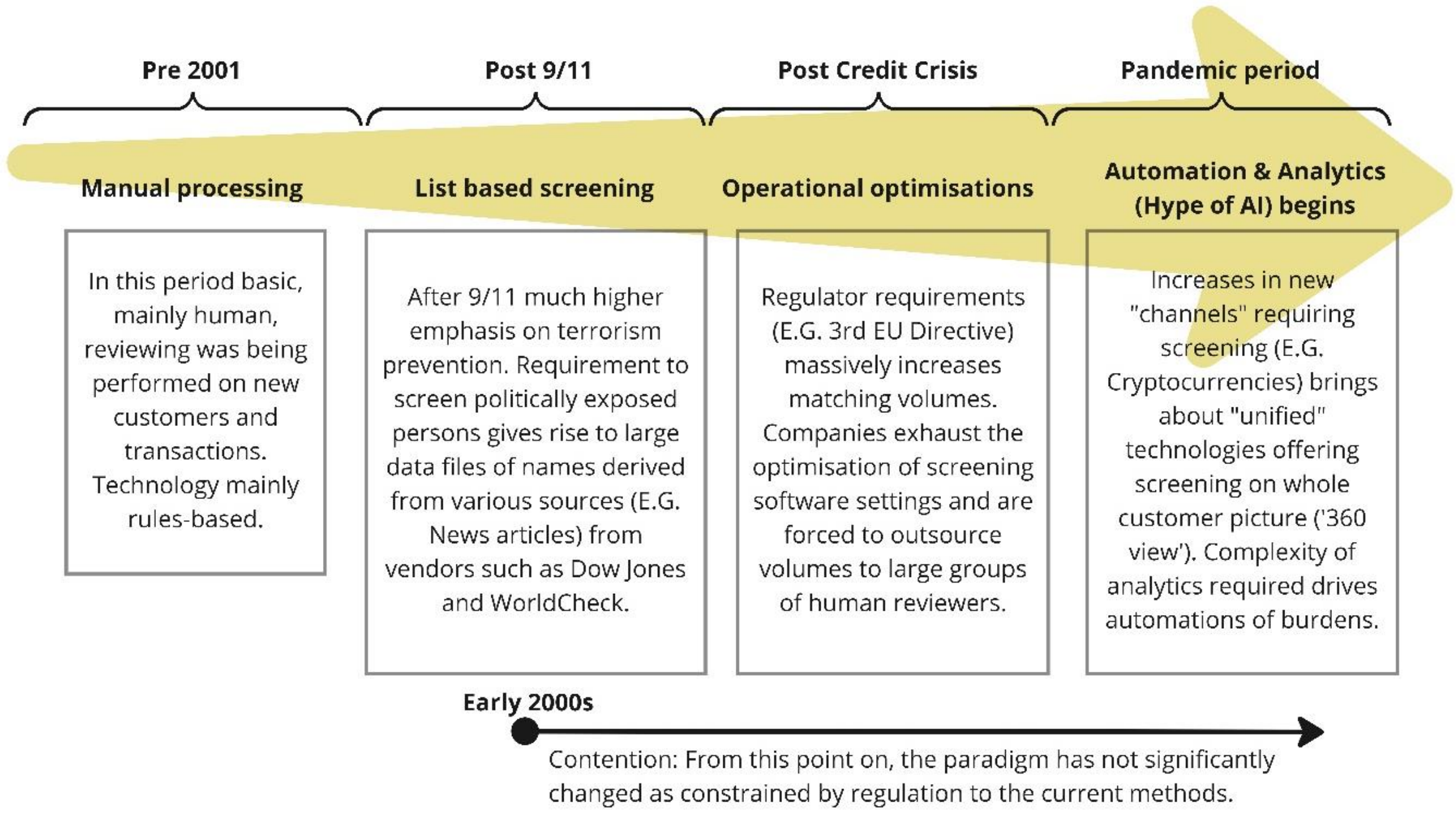

In this very mature technological arena, Artificial Intelligence offers relief through the prospect of disruptive innovation. As explained by McCaul in her speech to the ECB in 2022 (McCaul, 2022), the hope is that AI innovations offer "tremendous opportunities for banks". However, as noted by Bell in his 2021 keynote to the CI Fellows (Bell, 2021), "Technology tends toward the removal of burdens" and not straightforward innovations. Consequently, AI technologies are often initially aimed towards directly replacing the volume burdens and, with it, the staff involved. Bertrand provides two diagrams of before and after, with the first having multiple humans-in-the-loop and the latter being a model of automated inhumanity (Maxwell, Bertrand and Vamparys, 2020).

*Figure 2 – Process re-engineering - Two options: Full automation by AI or augmentation of current human-led processes.*

*Source: the author*

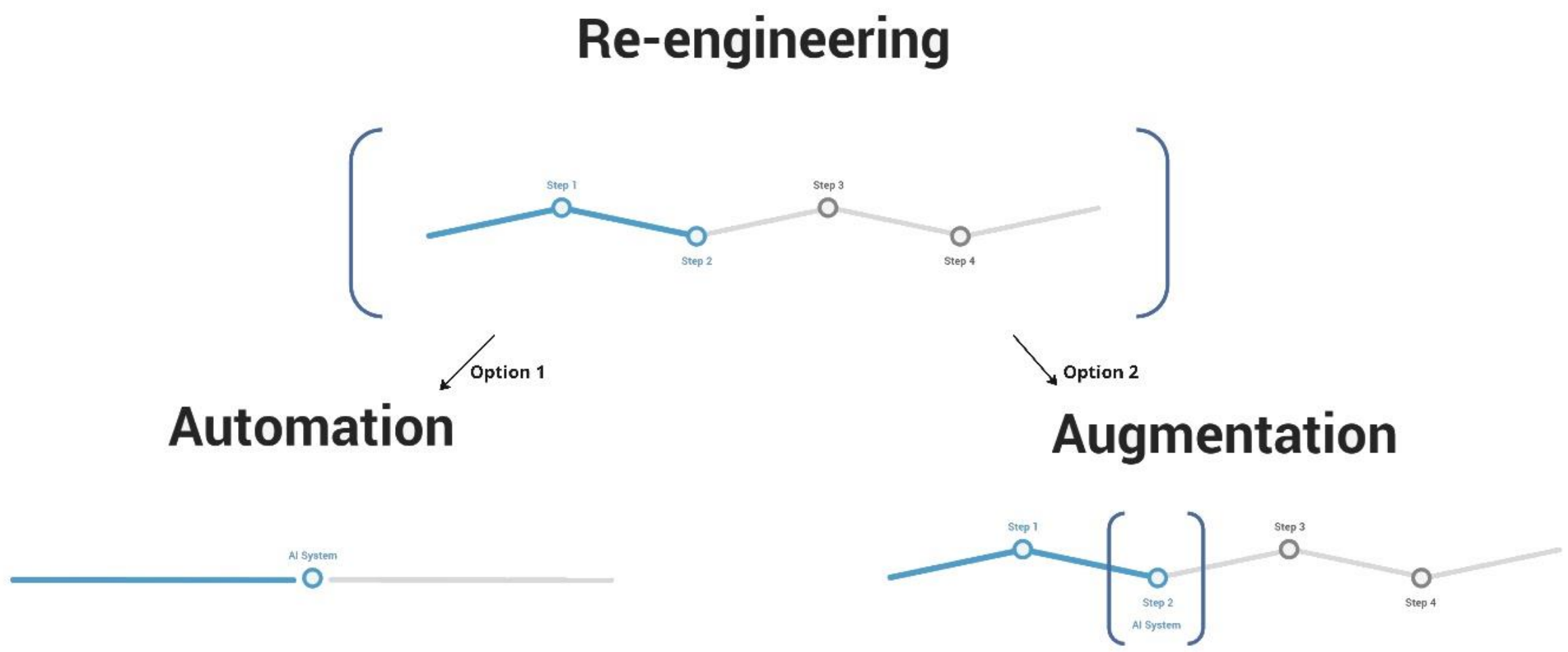

The economic impact of up to 95% of Citi compliance staff being laid off is uncalculated. That such AI technological disruption is possible is not clear because AI comes with problems for banks. As Kuiper notes (Kuiper *et al.*, 2022), explainability (the ability to trace an algorithmic system's thinking) is fundamental in many current and future regulations (such as the GDPR and upcoming EU AI Law (European Commission, 2022). A review of explainable AI systems used in AML by Kute (Kute *et al.*, 2021) concluded that only 49% of 43 current methods used in the industry are explainable. Consequently, banks may encounter problems concerning the fundamental human rights of the bank's customers and ethical concerns with automating hundreds of thousands of jobs out of existence.

The above literature review suggests the following testable hypothesis:

| | |
|---|---|
| **Hypothesis 1** | The risks of current-generation AI need to be better appreciated and understood by those working in the industry. |

**Article Review: The issue with SARs**

The wider economic benefits of transparency disclosures (in the form of SARs) to the 'UKFIU' was investigated by Tadesse in 2006 (Tadesse, 2006), with an examination of the impacts and consequences of transparent disclosure culture. This paper provides a broader lens as to why SARs are vital to the regulated sector. Furthermore, it illustrates an intractable problem; SARs are needed, but the regulatory grip creates too many to be helpful.

Tadesse demonstrates evidence that banks' transparency directly leads to a lowering of vulnerability to "crisis" states. Tadesse admits to the controversy of this relation, and it is apparent in the paper that the definition of "crisis" is critical to the proper evaluation. Tadesse uses the notion, formed first by Demirguc-Kunt and Detragiache (Demirguc-Kunt and Detragiache, 1998) that a crisis is where bank capital has been "exhausted".

Tadesse contends that competing forces are at work in regulating banks. One force, named "Transparency-Stability" (Tadesse, 2006, p. 33), suggests that transparency is a factor in market discipline, leading to better identification of weaker banks that risk damaging the banking system. The other opposing force, "Transparency-fragility", suggests that transparency can harm the banking system because it can highlight systemic faults and lead to crisis states.

These two forces are supposed to balance each other out. Indeed, SARs should identify a specific risk weakness with a regulated company or one of its customers and hence show the protective aspects of transparency to the whole industry. However, the torrent of SARs, to the point that the UKFIU fails to process them all effectively, demonstrates that excessive transparency instead highlights a systemic weakness in the regulation of the industry.

As such, while Tadesse's conclusion is not definitive, and their methods rely on tightly controlled data, identifying the core issues around transparency applies directly to this paper's aims and objectives and highlights the underlying volume problem well.

Tadesse’s work suggests the following testable hypothesis:

| | |
|---|---|
| **Hypothesis 2** | The regulators and governmental bodies act as "blocking", "protecting" or "safeguarding" devices stalling Innovation. |

**Book Review: A theoretical basis for change seen in the AML industry**

In order to produce a third testable hypotheses from which this paper's claims can be argued from the research, it is vital to examine how paradigms are replaced in detail through an analysis of the process of creative destruction (‘CD”), developed by Joseph A.Schumpeter in "Capitalism, Socialism and Democracy" (Schumpeter, 1976).This theoretical grounding will ensure proper consideration is given to suitable research methods, interpretation of the results, generalisability of the conclusions and predictions of future outcomes.

Two review questions are addressed:

1. To what extent does the economic theory of CD help explain the operation of the AML market?
2. Can this theory identify gaps in the general understanding of this market and highlight any risks inherent in the current paradigm?

Schumpeter explains the forces of market innovation using the lens of Marxian economic analysis. His goal with this research is to provide some contextual criticism of the standard Marxian economic models and then to use his insight to help explain a fundamental and inherent process of the capitalist economy.

Schumpeter agrees with Marx by identification of the exploitation inherent in a capitalist system (Schumpeter, 1976, p. 26). This identification is because innovation drives the labourer to work more hours than is required to produce stock for sale, resulting in a "surplus value" (Schumpeter, 1976, p. 27), which is then turned into capital with the value going to the capitalist. By a "compulsion to accumulate" (Schumpeter, 1976, p. 31) surplus value and thereby gain a higher market share, the capitalist fosters innovation within the company by devising new production methods and technologies. Schumpeter believes this is a constantly moving force threatening to upset any equilibrium in a market.

**The process of creative destruction**

Schumpeter likens this to an evolutionary process, which is constantly moving, and acts by industrial "mutation" to drive economic change (Schumpeter, 1976, p. 83). Schumpeter names this process "creative destruction" (Schumpeter, 1976, p. 81) and explains how it is often confused for the administration of existing structures. However, when witnessed acting over time, it is seen as part of the larger business cycle, busy creating and destroying dominant paradigms.

Creative destruction is a technical and not pricing-based revolutionary economic process, producing significant impacts on existing structures, and one where economic progress trumps any moral concerns. Schumpeter notes that certain "protecting" or "safeguarding" devices (for example, patents) (Schumpeter, 1976, p. 88) act in opposition to steady the market, at least enough to enable the more extensive forms of investment. However, he also notes that there is little to no point in preserving obsolete industries as long as there is no crash precipitating a depression.

Schumpeter notes that the moral dimension is often disregarded because the capitalist stands to gain from both an increase in efficiency and market share, and the whole process is seen as inevitable within the capitalist economy.

**Evaluation of Schumpeter's claims**

Schumpeter's explanation that worker exploitation is the driving force behind the CD process is readily identifiable within the similarly exploitative methods used to create AI technologies, where AIs are trained on historical worker "activity" data.

For example, in the case of AML reviewing, the human staff members decide on the status of a review match based on specific regulatory rules around suspicions. Replicating that decision-making ability with an AI requires that human "thinking" activities (such as searching third-party databases, reading the information in other records, and reviewing activity over time) are also captured as part of the decision data.

While the capture of the "doing" process has been typical in many manufacturing worlds, for example, the use of robotics in manufacturing cars, the more complex "thinking" activities of AML reviewing were thought to be too advanced to replicate. However, the rise of AIs suggests that a viable simulation of these thoughts can be achieved if enough captured thinking data is used as input into the algorithm's training. Therefore, the very thoughts of the workforce are exploited along with the products of their labour. Nevertheless, they receive only redundancy and reduced income prospects for training a technological simulation as a replacement.

Schumpeter's analysis of the operation of CD provides the theoretical economic context for the technological approaches to combat financial crime and allows for assessing the current AML paradigm. However, it also highlights the gap in current thinking, which risks economic progress and creates a pressure cooker effect.

Schumpeter provides the final testable research hypothesis:

| | |
|---|---|
| **Hypothesis 3** | The Anti Money Laundering industry is experiencing creative destruction through the pursuit of AI. |

## Chapter 3. Research design

### Introduction

The literature review chapter identified three hypotheses, the following methodological tests will be used to investigate and produce evidential research to support them:

- News topic time series using Factiva.
- Sentiment analysis using a Factiva extract.
- Questionnaire to regulated persons.
- Interviews with AML professionals.

Each will be now covered in detail with a summary at the end of how they answer the hypotheses.

### News data via Factiva

News data provides insight into what is considered publically important. News information acts as a data generator, measuring the social aspects of human thought filtered through editorial appraisal. News articles can be used as a single story pinpointing details on a subject or in the aggregate to clarify trends. This aggregate approach aligns with what Schumpeter identified (see literature review chapter) as the requirement to spot the CD process in action. News, therefore, meets the needs of all the identified objectives.

Over time the rise of a newsworthy keyword follows the concept of an "information lifecycle" adapted by Chun Wei Choo (Choo, 2007). This proposed that topics (a combination of keywords) have gone through several "weak signal" stages until appearing in the mainstream press. Once in the mainstream, further news articles "hype" the topic until it reaches the attention of governments, which inevitably regulate it.

*Figure 3 - Choo, C. W. (2007). Information life cycle of emerging issues. (Adapted from (Wygant and Markley, 1988))*

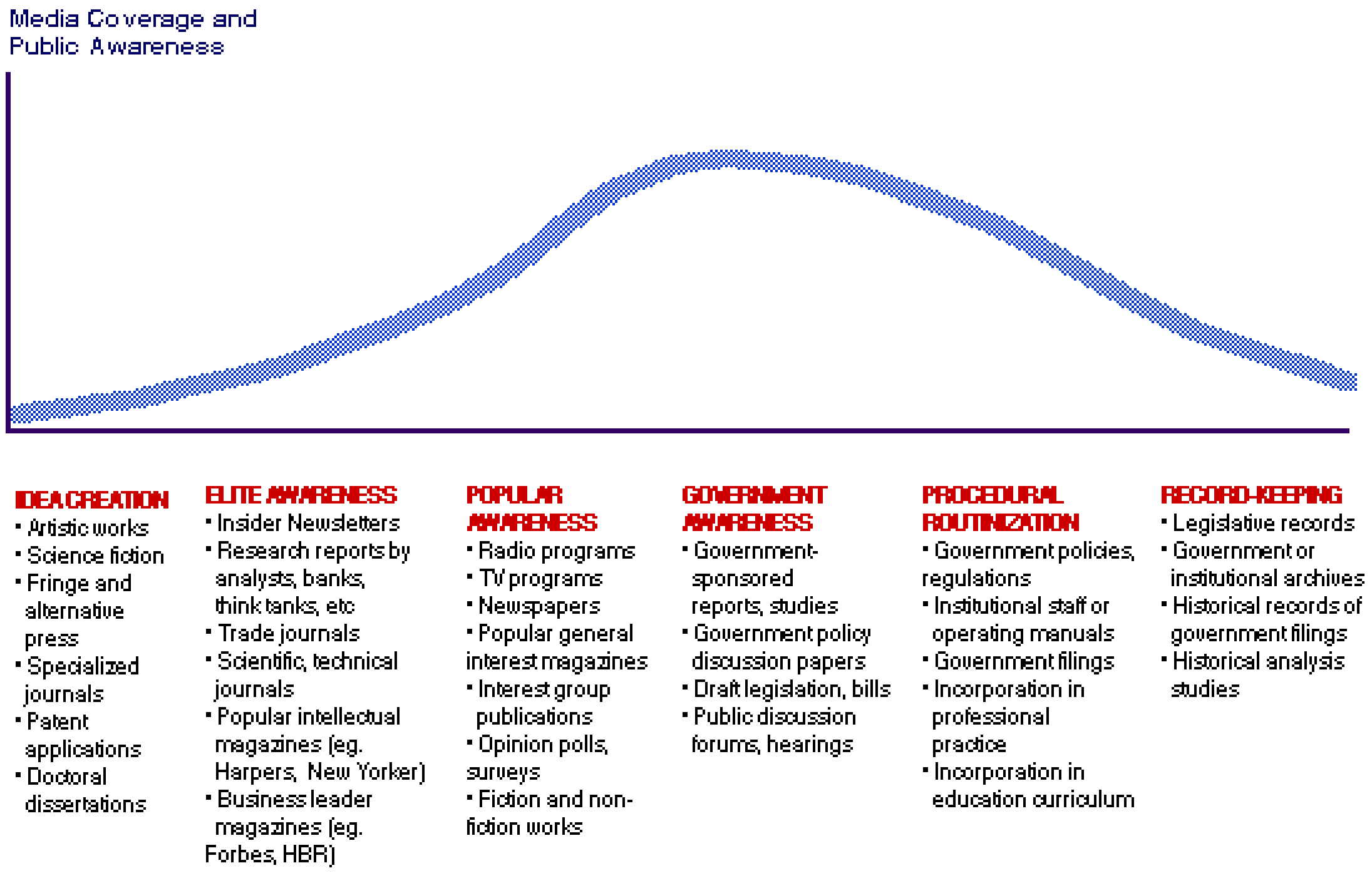

Factiva provides a repository of newspaper articles with over 1.6bn pieces from 8,900 sources available as API downloads through the analytics service. Searches made in Factiva can be directly read and reviewed to answer hypothetical questions. In addition, a summarisation of results will be available to provide primary statistic time series, such as volumes of articles with specific keywords arrayed over time.

### Factiva aggregate search settings

A search was undertaken to plot the volume of articles on the subject of “AI” and “ML”. The exact search settings were:

| | |
|---|---|
| **All of these words** | "Artificial intelligence" |
| **At least one of these words** | "money laundering" |
| **Date*** | In the last 10 years |
| **Source** | All Sources |
| **Author** | All Authors |
| **Company** | All Companies |
| **Subject** | All Subjects |
| **Industry** | All Industries |
| **Region** | All Regions |
| **Language** | All Languages |
| **News Filters** | Subject: NOT Press Releases \| Source: NOT Securities and Exchange Commission (SEC) Filings (U.S.) |

The news filter removed press releases and SEC filings from the results as they are irrelevant to general opinion.

**Factiva sentiment search settings**

Sentiment analysis will be completed on a randomised sample size of 200 articles by reading the samples directly and scoring based on the following scale:

1. The article is very negative about AI in AML, mentioning specific issues or highlighting risks. May specifically mention job losses.
2. The article includes negative aspects of using AI in AML and more general points.
3. The article discusses or mentions AI in AML but does not specify either benefits or drawbacks of the technology.
4. The article is generally positive about the use of AI in AML and mentions only its potential benefits or highlights a production use case.
5. The article is very positive about AI in AML and hypes the potential of the technology without much regard to evidence of the benefits being realised.

The exact Factiva sentiment search settings were:

| | |
|---|---|
| **All of these words** | "Artificial intelligence" |
| **At least one of these words** | "money laundering" |
| **Date*** | In the last 2 years |
| **Source** | All Sources |
| **Author** | All Authors |
| **Company** | All Companies |
| **Subject** | All Subjects |
| **Industry** | All Industries |
| **Region** | All Regions |
| **Language** | All Languages |
| **News Filters** | Subject: NOT Press Releases \| Source: NOT Securities and Exchange Commission (SEC) Filings (U.S.) |

*Figure 4 - An example Factiva Result, note how the system highlights the search keywords. Source: Factiva.com*

Future Of Finance
**Only 1% of laundered cash in EU is detected — ABN AMRO wants to improve that; ABN AMRO's Ivich Hoffman reckons detecting money laundering is like mixing the perfect cocktail**

David Canellis
840 words
17 November 2020
The Next Web
NEXTWEB
English

Technology has simplified the movement of money so much that people and businesses can send funds anywhere in the world, instantly, but tech's rapid advance has also brought new ways to launder cash.

In fact, the United Nations warns this rise of "megabyte money" makes curbing the transfer of illicit funds more urgent than ever.

Globally, estimates suggest between $800 billion and $2 trillion in laundered money flows through the financial system every year, and an overwhelming majority of it goes undetected. The Netherlands alone sees $16 billion in criminal money flowing through its financial system.

Zooming out to Europe as a whole, the European Commission found just 1% of an estimated $190 billion in laundered funds were successfully confiscated between 2010 and 2014.

To help uncover **money laundering** across the EU, ABN AMRO teamed up with ING, Rabobank, Triodos Bank, and de Volksbank to share all of their customers' payment transactions in an initiative creatively dubbed (TMNL).

Indeed, thanks to a new Dutch law allowing banks to share limited customer data, 12 billion transactions per

Factiva has limitations because not all possible sources are available in the repository. This limitation is why this paper also uses direct questionnaires (discussed below). Similarly, news data can contain bias. However, since this project is about detecting that bias, as a reflection of the underlying positive sentiment surrounding AI and AML highlighted in the literature review, this is considered to have a positive impact.

Ethically, using Factiva ensures that premium news data has been used under its license for research purposes. Aggregate stats also ensure individual journalists are privacy protected.

### Participant selection methods

The selection criterion focuses on industry insiders, firstly because the complex nature of AML lends itself to understanding only by those involved in the industry. Laypeople struggle to comprehend the highly technical nature of AML practice because the POCA demands a veil of secrecy in its operation to prevent tipping off. Secondly, these insiders operate under the regulated burden and are, therefore, the opinions required.

The following sources provide the audience setting for questionnaires:

1. The author's network. These individuals will be contacted directly using email addresses provided on business cards.
2. LinkedIn searches for AML professionals with 1st-level connections to the author. LinkedIn enables quick and easy connections to professionals based on their credentials. This group encompassed those in the author's network whom he has directly worked with, even at a distance, and provides a wide spread of AML professionals globally. This group encompassed a potential audience of 326 people.
3. 2nd-level connections to the author. The search was limited to persons with "MLRO" in their job history within the last two jobs. This filter was selected because it includes those who have left the MLRO role (for example, retired or promoted to another role). This method provided a higher chance that the contacted individual would respond. Mainly because of the connection to a third person who was a 1st-level connection to the author. This group encompassed a potential audience of 3,300 people. LinkedIn allowed connection to 500 in the time the project allotted for this task.
4. "Followers" of the author on LinkedIn; a much wider group pool than AML professionals, including a large pool of AI professionals. This method provided for a much larger "self-selection" of people than those directly contacted, and so the questionnaire was designed to filter out the responses that might not be relevant. This group encompassed a potential audience of 3,317 people.

The post the author placed on this subject had 2,511 impressions (The estimated total number of times the post was displayed on a screen) with a demographic breakdown as follows:

*Figure 5 - screenshot of LinkedIn post analytics*

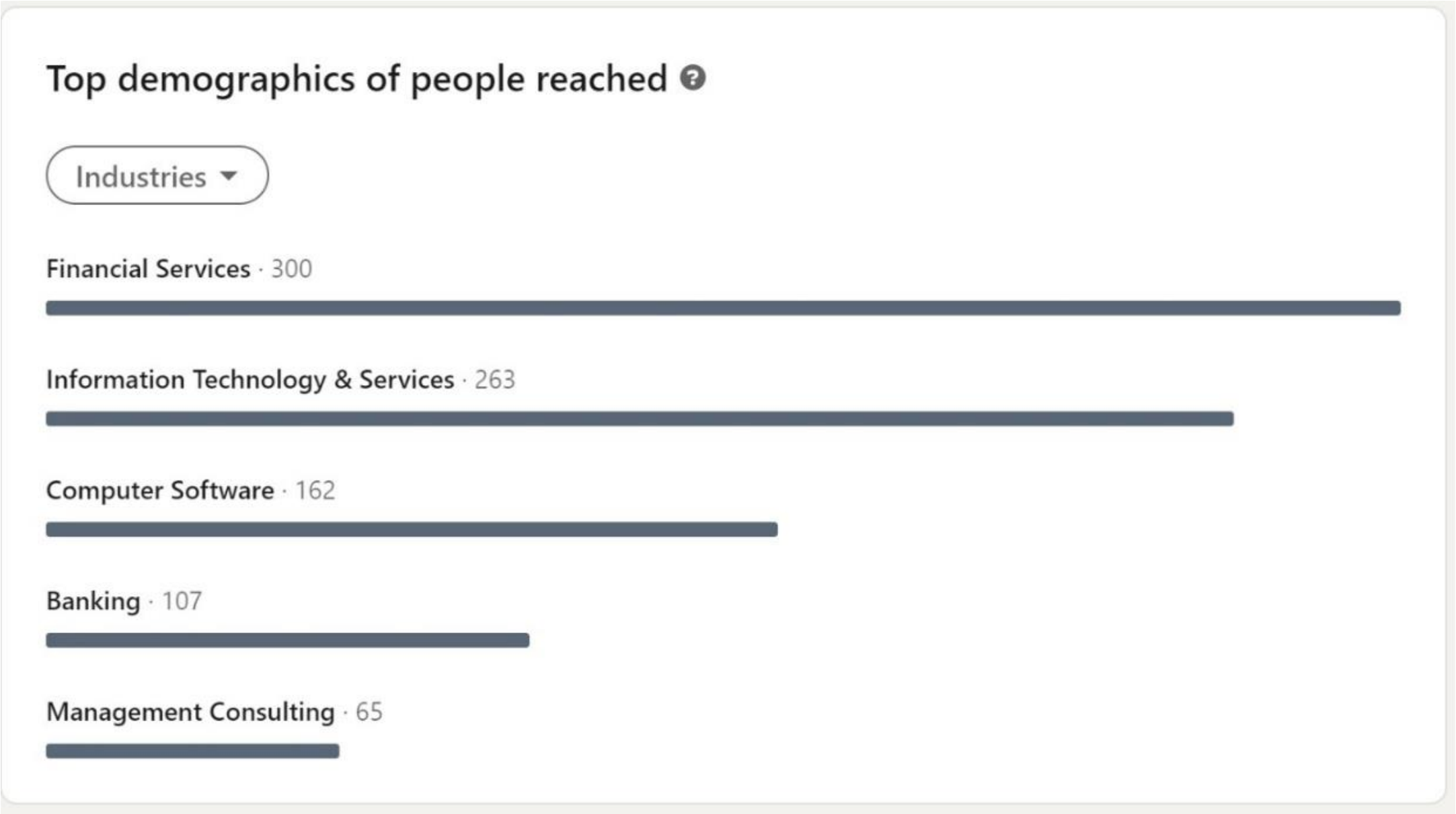

One drawback of using LinkedIn in this way is the possibility of the author's opinions having affected the participant’s thinking around AML due to their career connections. This issue is why the group pool extended to 2nd-level connections. Both to widen the pool beyond whom the author has met and tap into the network effect of connection relevance. This effect ensures a higher quality of connections in the group pool and the relevance of their opinions.

The primary ethical consideration of using LinkedIn is the question of anonymity. Therefore the questionnaire will need to ensure that personally identifiable data is not captured in its design.

**Questionnaire**

The questionnaire setting is surveymonkey.com. This site was selected because it provides the following benefits:

1. Cost reduction through University email: a discount on a professional account.
2. One link serves participants an identical experience.
3. Open, closed and optional questions increase the survey's completion rates and engagement.
4. Feedback on the questionnaire directly.
5. Good integration with LinkedIn, which increases take-up.

**Sample size**

The minimum number of respondents was calculated to increase the generalisability of the research results. The population size of the UK "AML population" was detailed in the "Anti-money laundering and counterterrorist financing: Supervision Report 2019-20" by HM Treasury (HMTreasury, 2021, p. 10). This report notes that the size of the UK "AML population" is 19,660. This number would include some roles not relevant to the paper.

This finite population gives the following results when calculating the minimum population size for the sample size:

$$\text{Finite population:} \quad n' = \frac{n}{1 + \frac{z^2 \times \hat{p}(1-\hat{p})}{\varepsilon^2 N}}$$

where

**z** is the z score

**ε** is the margin of error

**N** is the population size

**p̂** is the population proportion

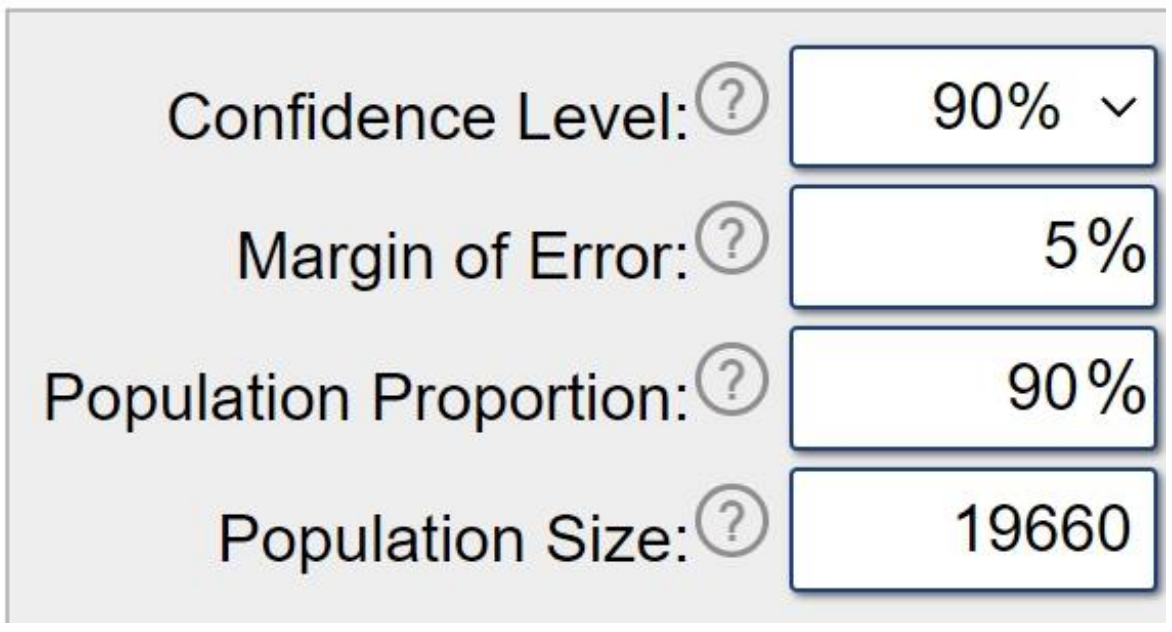

**Result Sample size: 98**

Because of this sample size requirement, the questionnaire will require at least 100 respondents. By having a statistically relevant population sample size, the questionnaire answers objectives 1.

Questions allow for a blank response to filter irrelevant opinions to the research topic.

**Interviews**

The selected setting for interviews is "semi-structured", as detailed in chapter 10 of "Research Methods for Business Students" (Saunders, Lewis and Thornhill, 2019). Interviews will use themes developed from the questionnaire, which enables an in-depth investigation of points and an opportunity for any contradictions to be clarified. The responses to open-ended questions will be of particular interest, which can highlight deeply held attitudes and provide significance to the questionnaire results, thus meeting the requirements of objectives 1 & 2.

Interviews are either:

1. In-person. The interviewee will select a meeting place, and the author will travel to the location to conduct the interview.

2. Google Hangouts at a mutually acceptable time (given time zone constraints,) and the author will send a "hang out request" via email.

In both instances, phone-based recordings will be summarised into notes. Interviews are to complete within an hour.

Semi-structured interviews can contain bias so consideration will be given to the appropriateness of the interview location and the opening remarks framing the interview.

Ethical considerations for interviews are many. Recordings of unfiltered responses "off-the-cuff" can impact a person's career if heard elsewhere. Therefore, the recordings will be stored on a secure removable thumb drive and encrypted.

An information sheet outlining their rights and expectations is given to each interviewee to ensure they know that the meeting is recorded. Selected quotations will only be used anonymously and with all gendered terms and company names removed.

**Questionnaire design**

The objectives and related testable hypotheses (H1-3) align with the questionnaire as follows:

| Section | No. | Question | Type | Cross-linked | Objective | Hypothesis | Reason / Notes |
|---|---|---|---|---|---|---|---|
| A | 1 | Which of the following best describes your current occupation? | Closed / Ordinal | All | N/A | N/A | Important for categorisation of later sections |
| A | 2 | Which of the following best describes your current job level? | Closed / Ordinal | All | N/A | N/A | Important for categorisation of later sections |
| A | 3 | What is your age? | Closed / Ordinal | All | N/A | N/A | Important for categorisation of later sections |
| A | 4 | In what country do you work? | Closed / Categorical | All | Bio & Geo data | N/A | Important for categorisation of later sections |
| B | 5 | "Regulations significantly burden companies and those involved in combatting financial crime." | Closed / Ordinal | Q6 | O1 | H3 | This question specifically asks if the regulations are burdensome. Language, specifically "burden", directly from gov.uk in their "Action Plan for anti-money laundering and counter-terrorist finance" (HMTreasury, 2016). Placing it here enables cross-tabulation with later questions. |
| B | 6 | Do you think the regulatory burden has changed significantly in the last five years? | Closed / Ordinal | Q5 | O1 | H3 | This question asks, have regulatory burdens been getting worse or better? |
| B | 7 | If so, why? | Open | Q6 | O1 | H3 | This open question provides a space for their wider thoughts and ensures themes not covered in the prior question's responses are picked up |
| B | 8 | "AML methodologies have significantly changed for many years." | Closed / Ordinal | Q5 | O1 | H2 | Does the respondent think that methods to deal with changing regulations have also changed? |
| B | 9 | "The volume or rate of Suspicious Activity Reports has increased significantly since I entered the business." | Closed / Ordinal | Q3, Q2, Q4 | O1 | H3 | Have these methodological changes proved effective? |
| B | 10 | The amount of review matches in AML systems requires significant time to manage. | Closed / Ordinal | Q5, Q16, Q13 | O2 | H3 | Has your time spent reviewing AML matches increased? |
| B | 11 | If you are involved with reviewing matches: what proportion of time is spent reviewing matches in your organisation or organisations you have dealt with? | Closed / Numeral | Q10 | O2 | H3 | What is the match review burden for you? |
| B | 12 | Have you required the use of outsourced or offshore staff to handle volumes of matches? | Closed / Categorical | Q11 | O2 | H3 | Have you explored outsourcing for this burden? |
| B | 13 | "I think a more balanced approach to regulation and AML would be more productive to combatting financial crime." | Closed / Ordinal | Q5 | O2 | H3 | Do you think the regulatory burden is too high and restrictive? |
| B | 14 | The FCA said in 2019 that money laundering causes "incalculable" | Closed / Ordinal | Q13, Q6 | O2 | H3 | Do you agree that this burden is important? |
| | | damage to society and that technology might be the answer. Do you agree? | | | | | |
| B | 15 | Do you think the skills required to perform AML successfully are being lost due to the high burdens of the regulations? | Closed / Ordinal | Q14 | O2 | H3 | Does the regulatory burden impact the desire to work in the sector? |
| C | 16 | "I would say that AML technology has kept pace with regulation changes." | Closed / Ordinal | Q6 | O3 | H3 | Is technology change responding to regulatory changes? |
| C | 17 | "I would say that new and revolutionary technology is available to those in Anti-Money Laundering". | Closed / Ordinal | Q6, Q15 | O3 | H2 | Does AML technology have a revolution/generational shift/Paradigm shift? |
| C | 18 | Have you used AI technologies to help with the regulatory AML burden? | Closed / Categorical | Q17 | O3 | H2 | Is Artificial Intelligence one of the new technologies? Has the respondent used this? Are attitudes different between groups? |
| C | 19 | Do you think that these technologies represent good value for money to the business? | Closed / Ordinal | Q18 | O3 | H2 | Is AI technology expensive for the value it brings? |
| C | 20 | Have AI technologies improved your business's ability to fight financial crime? | Closed / Ordinal | Q18 | O3 | H1 | Is AI technology effective in reducing the respondent's regulatory burden? |
| C | 21 | Do you think AI can solve AML volume issues (or alleviate them)? | Closed / Ordinal | Q18 | O3 | H1 | Is it removing the problem or only alleviating it? |
| C | 22 | What are your opinions of the current maturity level of AI technologies in AML? (Question uses Technology readiness levels (TRLs)) | Closed / Ordinal | Q18 | O3 | H1 | Is the AI technology in production or another IT readiness state? |
| C | 23 | Do you think that AI for AML is meeting the industry's expectations? | Closed / Ordinal | Q18, Q5 | O3 | H1 | Does the respondent feel that AI technology is as good as it needs to be? |
| C | 24 | Do you think AML technology vendors successfully integrate AI into their products? | Closed / Ordinal | Q18, Q5 | O3 | H2 | Are AML technology vendors simply putting "AI inside" on the box? |
| C | 25 | Have you made use of, or heard of, "AI Ethics" | Closed / Categorical | Q18, Q1 | O3 | H1 | Are the respondent's organisation AI maturity levels high enough to consider AI ethics? |
| C | 26 | Do you think that AI for AML will lead to job losses? | Closed / Ordinal | Q18, Q1 | O3 | H1 | If the respondent has heard of AI ethics, have they considered the impact on jobs? |
| D | 27 | Do you have a few more moments to answer a few extra questions, leave some feedback or perhaps connect for a 1-1 interview? | N/A | N/A | | N/A | The final section is "extra voluntary" questions that only engaged respondents will answer. This section also sets up the interviews. |
| D | 28 | What does artificial intelligence ('AI') mean to you? | Open | Q18, Q1 | O3 | H1 | What is the respondent's understanding of AI |
| D | 29 | How do you perceive AI in the Market? | Open | Q18, Q1 | O3 | H1 | What is the respondent's understanding of the AI market? |
| D | 30 | Academia suggests a strong correlation between a bank's benefit to the economy and its ability to combat money laundering. Do you agree? | Closed / Ordinal | Q5, Q16, Q13, Q2 | O2 | H3 | Here referring to the Masciandro Model (Masciandaro, 1999) |
| D | 31 | "Crying wolf theory" states that defensive reporting "deluges" investigative units to the point that they cannot effectively act upon them. Do you agree? | Closed / Ordinal | Q5, Q16, Q13, Q3 | O2 | H3 | Are the regulated responding defensively to overlapping requirements from regulators and pressure from customers? |
| D | 32 | How would you rate this questionnaire? | Closed / Ordinal | N/A | N/A | N/A | Questionnaire review |
| D | 33 | Do you have any comments, questions, or concerns? | Open | N/A | N/A | N/A | General feedback |
| D | 34 | Would you be interested in a one-to-one interview? | Closed / Categorical | N/A | N/A | N/A | Setting up interviews |
| D | 35 | Contact details | Open | N/A | N/A | N/A | Details for interview |

**Synthesis of testable hypotheses**

The claims in this paper will be tested as follows:

| No. | Testable hypothesis | Methodological Tests | Objectives |
|---|---|---|---|
| H1 | The risks of current-generation AI need to be better appreciated and understood by those working in the industry. | Factiva Data Collection and Sentiment Analysis shows high positive bias. | 1 (Substantive) |
| | | The questionnaire replies demonstrate a highly positive response to "AI and AML" technologies, while at the same time showing concerns with its effectiveness as a solution | 2 (Substantive) & 3 (Theoretical) |
| | | Interviews highlight a high positive sentiment of AI technologies in general and in AML in particular | 3 (Theoretical) |
| H2 | The regulators and governmental bodies act as "blocking", "protecting" or "safeguarding" devices stalling Innovation. | Questionnaire responses show significant evidence of regulation's impacts placing an increasing burden on the regulated. | 3 (Theoretical) |
| | | Interviews highlight that the AML regulatory "burden" is high and/or the volumes are causing struggles. | 2 (Substantive) |
| | | Interviewees comment on the high levels of SARs required to remain within the regulations. | 2 (Theoretical) |
| H3 | The Anti Money Laundering industry is experiencing creative destruction through the pursuit of AI. | Factiva Data Collection shows an increase in "AI and AML" mentions in the news. | 1 (Substantive) |
| | | Questionnaire responses show significant evidence of the increasing use of AI in AML. | 1 & 2 (Substantive) & 3 (Theoretical) |
| | | Interviews mention the high relevance of AI technologies being released/used in the AML sector. | 1 & 2 (Substantive) & 3 (Theoretical) |

Evidence and answers to these tests will provide the bulk of the analysis results.

## Chapter 4. Findings and Analysis

### Factiva research findings

This research method considers the topic's information life cycle to pinpoint how "hyped" and "mature" the topic is.

The ten years of Factiva news data was aggregated into a time series using the analytics functions of Factiva as follows:

*Figure 6 - Factiva analysis of AI and AML search terms over 10 years*

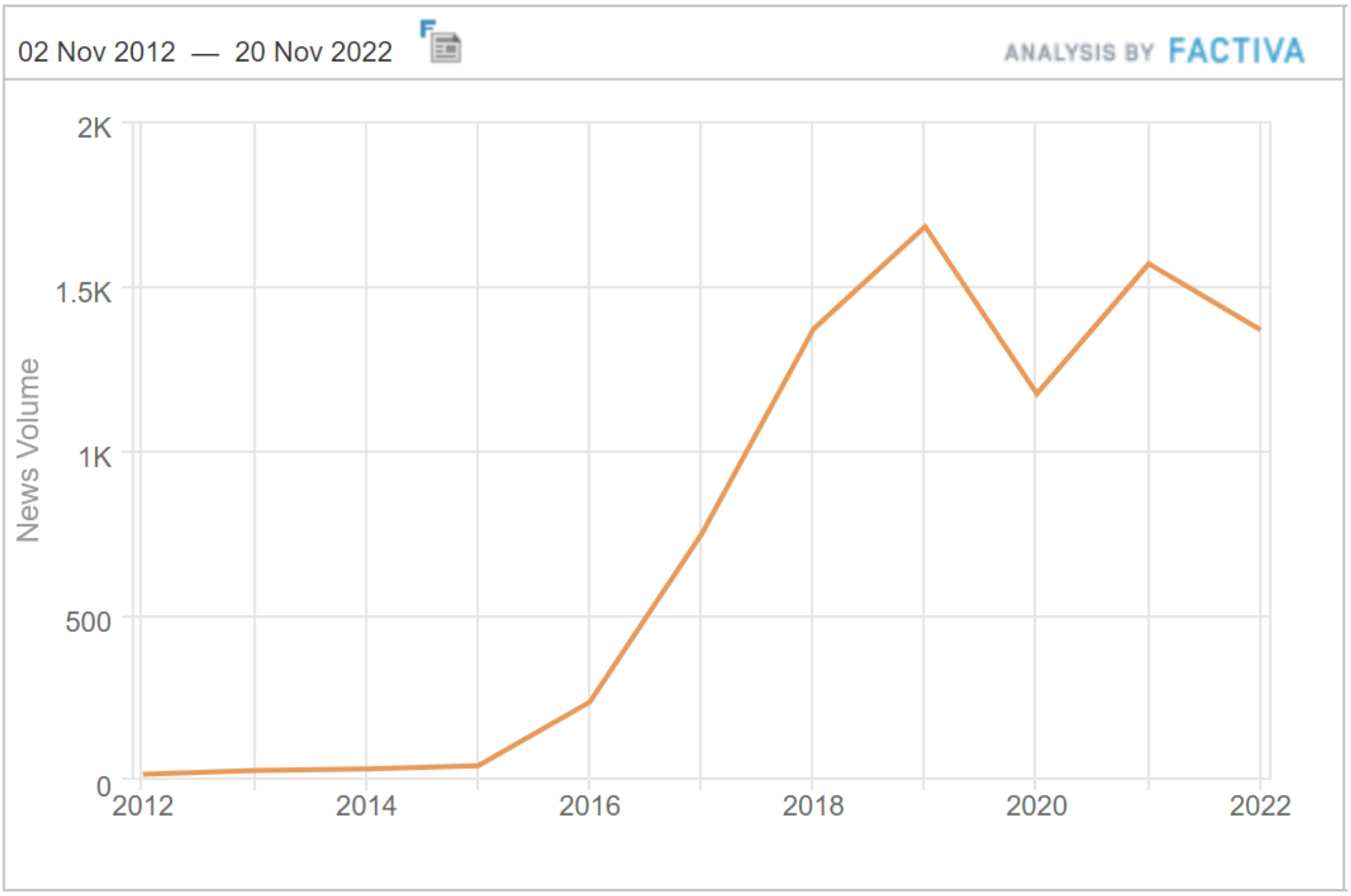

| Date | 2012 | 2013 | 2014 | 2015 | 2016 | 2017 | 2018 | 2019 | 2020 | 2021 | 2022 |
|---|---|---|---|---|---|---|---|---|---|---|---|
| Article Count | 23 | 29 | 34 | 43 | 237 | 747 | 1,369 | 1,680 | 1,172 | 1,568 | 1,281 |

This chart shows that the topic for "AI" and "ML" has grown and increased in line with the Information Lifecycle theory predictions outlined in the Research Design section. Further supporting evidence can be gained from analysis of the sources of news.

Most mentioned sources:

*Figure 7 - Sources from Factiva analysis*

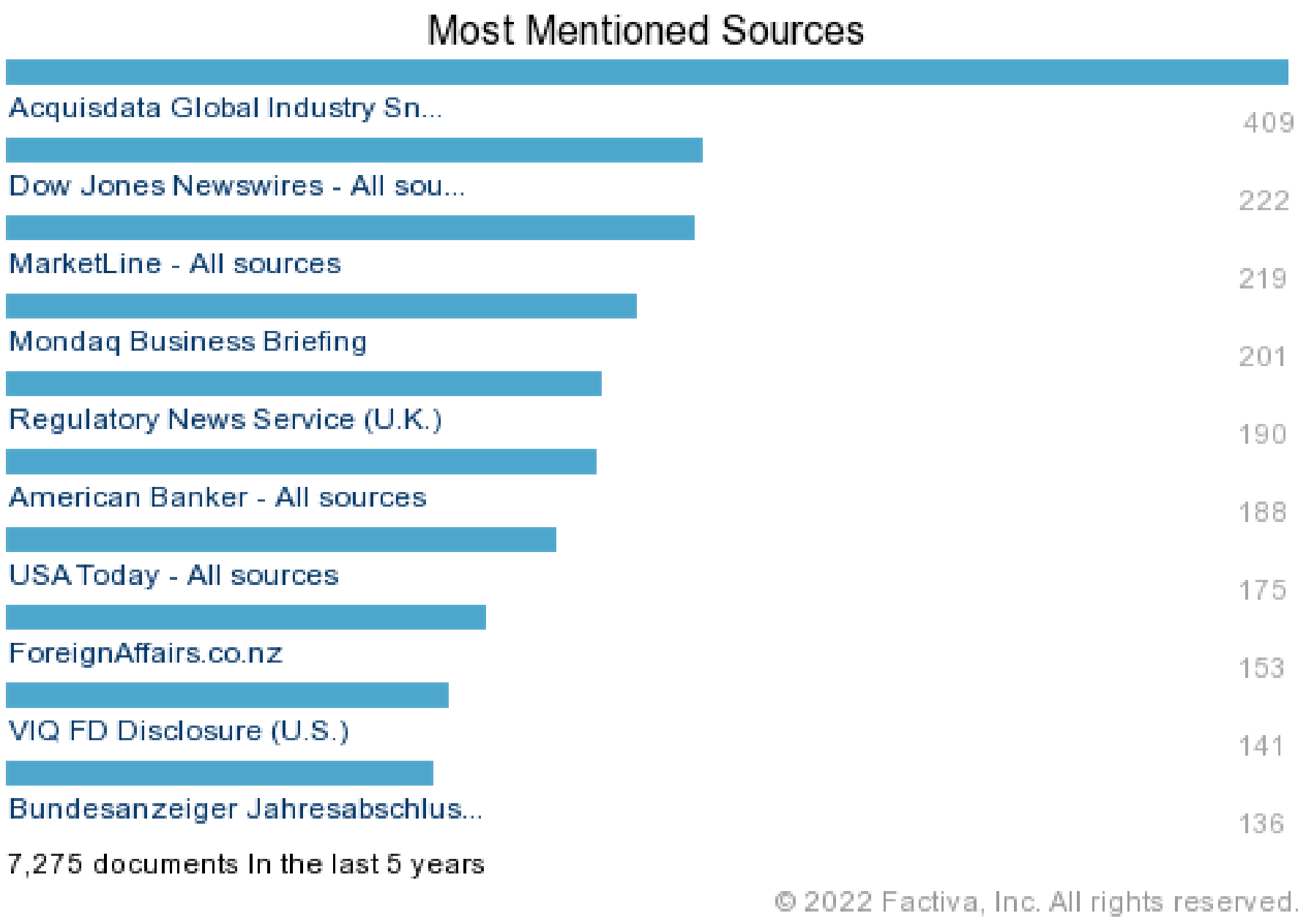

This chart includes many “mainstream” news sources, such as Newswires and USA Today. This suggests that the topic has left the “weak signals” stage and is following the lifecycle to reach a peak of hype. It should also be noted that the theory predicts that the topic will wane considerably in the later lifecycle (Choo, 2007). This waning is because the cycle begins anew with a new topic set to replace the current one. This follows the theory of CD and synthesises well with the lifecycle concept.

Similarly, the results included a wide range of global regions, which suggests that this phenomenon is global in scope, something that will be reinforced in the questionnaire and interview results:

*Figure 8 - Regions from Factiva analysis*

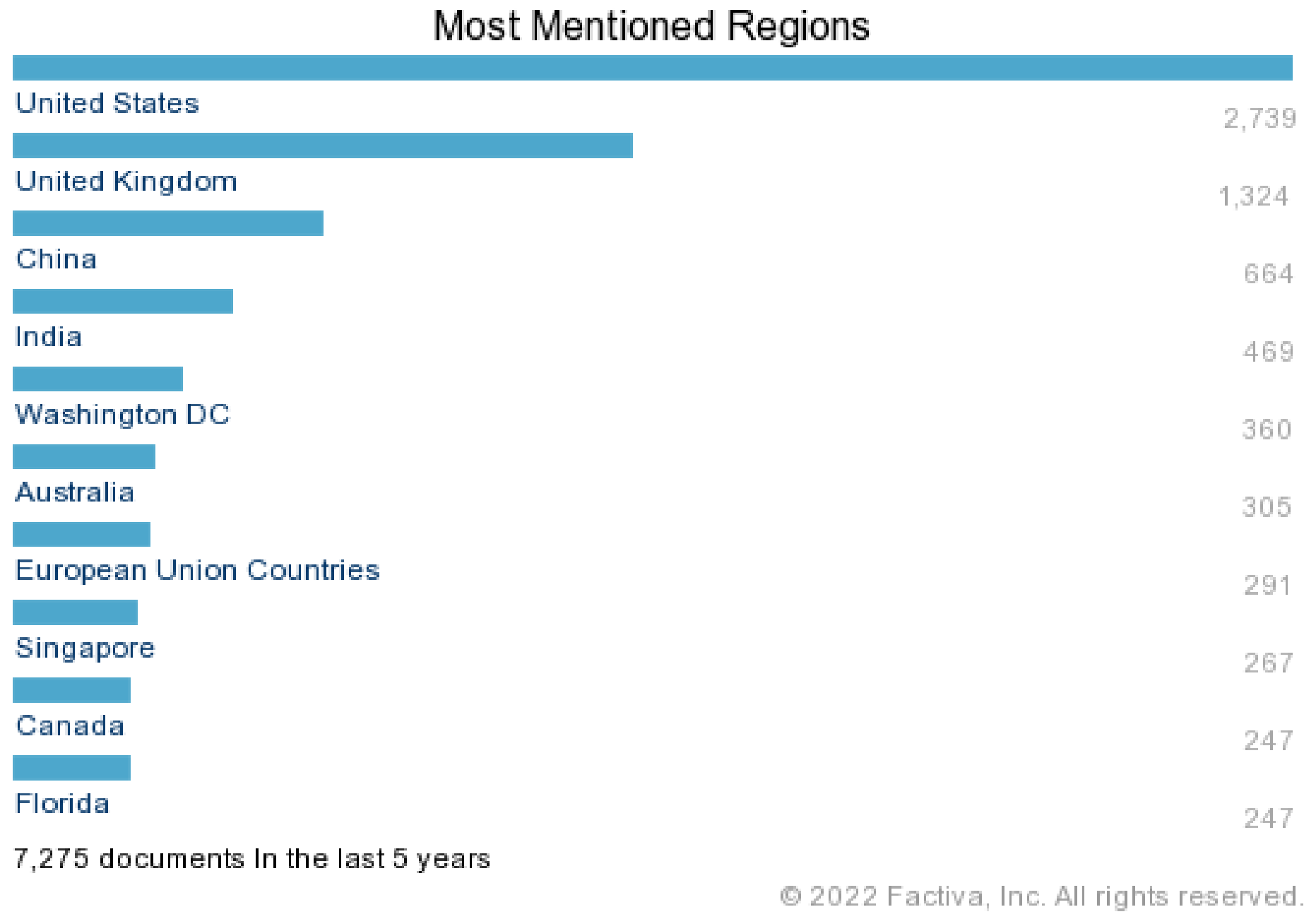

The topic also demonstrates a broad exposure in terms of the regulated industries covered in the news articles:

*Figure 9 - Industries from Factiva Analysis*

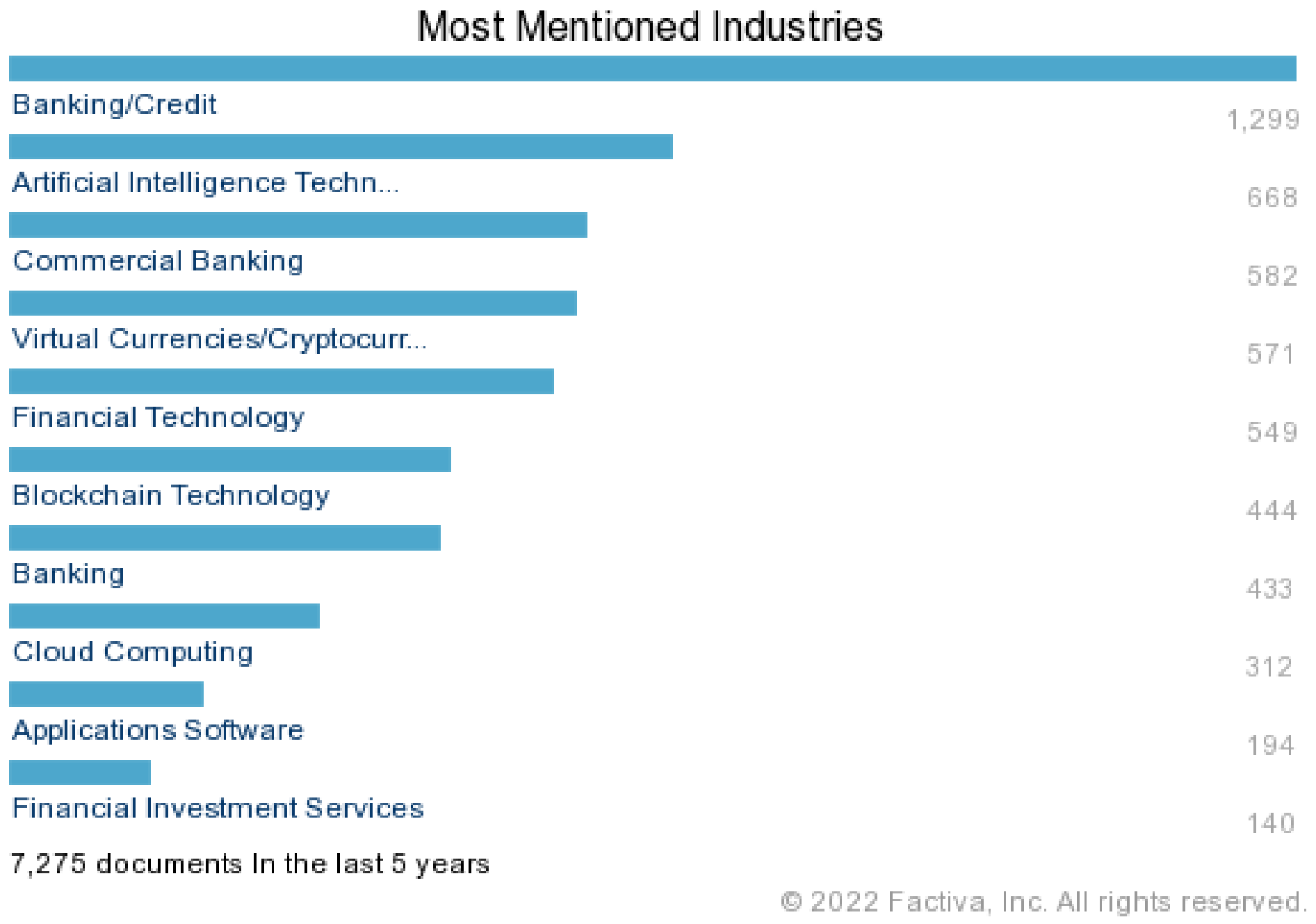

These results suggest that the Factiva research supports Hypothesis 1.

**News article sentiment analysis**

Sentiment analysis on the Factiva results considers how positive each article is about “AI” and “ML” to illuminate any bias.

The search was filtered to only the prior two years, and random articles were manually reviewed and scored. Results have been ordered by rating:

| Article | Sentiment Rating |
|---|---|
| US sought records on Binance CEO for crypto money laundering probe | 1 |
| Best weapon against money laundering is people, not tech – global watchdog | 1 |
| UN counter-terrorism body backs innovations to fight digital terror | 1 |
| UNOG - UN counter-terrorism body backs innovations to fight digital terror | 1 |
| AI use carries bias risk for financial regulators | 1 |
| Binance comes out swinging as report links it to money laundering | 2 |
| IEU MONITORING for 07/20/2021 | 2 |
| Regulation, the biggest challenge to apply anti-money laundering technologies: FATF study | 2 |
| Danish reg’ rolls out two-year fintech plan | 2 |
| ACAMS' 12th Annual AML & Anti-Financial Crime Conference -- APAC Takes Cutting-Edge Look at the Perils and Promise of Compliance Efforts in... | 2 |
| MONEY LAUNDERING; private sector wants European regulation to be aligned with international standards | 2 |
| Europol: better equipped to fight crime and terrorism | 2 |
| Teleperformance SE - Banks Need a New Approach to Fight Money Laundering | 2 |
| Outrage as banks shut customers' accounts and freeze payments as small as £50 in panic over fraud threat | 2 |
| Regulation, the biggest challenge to apply anti-money laundering technologies: FATF study | 2 |
| After the presentation of a proposal for a European regulation of Artificial Intelligence What are the benefits for the gaming industry? | 2 |
| Review, Then Reform? AMLA Charts A Path For The Future Of SARs And CTRs | 2 |
| MIL-OSI Europe: Europe must act as a single country against organised crime | 2 |
| How banks use AI to catch criminals and detect bias | 2 |
| Bank money laundering efforts increasingly impact on ordinary account holders | 2 |
| The Role Of The UK Financial Conduct Authority In A Changing Regulatory Landscape - Speech By Nikhil Rathi, FCA Chief Executive, Delivere... | 2 |
| Bank of England - Machine learning in UK financial services | 2 |
| Gambling giant Entain could lose UK licence after record £17m fine | 3 |
| Digital Economy Drives Growth | 3 |
| Insight EU monitoring - Your Daily Premium Digest | 3 |
| Anti-Money Laundering Software Market Forecast to 2028 - COVID-19 Impact and Global Analysis By Component (Software and Services... | 3 |
| MasterCard International Inc Files Patent Application for Anti-Money Laundering Methods and Systems for Predicting Suspicious Transactions Using Artifical Intelligence | 3 |
| Banks can expect return to strict oversight of anti-money laundering rules | 3 |
| Money laundering will be too easy in the metaverse | 3 |
| Money laundering will be far too easy in the metaverse | 3 |
| Mastercard International Incorporated; Patent Application Titled "Anti-Money Laundering Methods And Systems For Predicting Suspicious Transactions Using Artifical Intelligence... | 3 |
| Insight EU Future of Finance Monitoring - January 2022 | 3 |
| Transaction Monitoring Market - Growth, Trends, COVID-19 Impact, and Forecasts (2022 - 2027) | 3 |
| Global Anti-Money Laundering Solutions Market - Growth, Trends, COVID-19 Impact, and Forecasts (2021 - 2026) | 3 |
| China central bank to establish crypto asset transaction tracking system | 3 |
| INDUSTRY SNAPSHOTS; Australia Tourism 11 Oct 21 | 3 |
| EDITORIAL / Financial loopholes must be closed to clean up money laundering | 3 |
| IDBI Intech Implements Anti Money Laundering Reporting Solution At LIC | 3 |
| Japan warned to boost measures against money laundering | 3 |
| This Year's Lloyd's Startup Accelerator Program Focuses on Flood Loss, Severe Weather and More | 3 |
| Lloyd's Selects 11 Start-Ups for Innovation Lab Accelerator | 3 |
| Payment Monitoring Market To Grow Extensively at Unstoppable Rate \| ACI Worldwide, BAE Systems, Fiserv | 3 |
| NICE Actimize; NICE Actimize Named "Best Compliance" Technology Provider by 2021 WealthBriefing European Awards | 3 |
| ThetaRay turns to Google Cloud to help reduce cross-border payment risk | 3 |
| SIM registration essential to Know Your Customer data centralisation- BoG | 3 |
| Video-identification and its legal evidentiary validity | 3 |
| Government must make sure regulation of fintech does not stifle innovation in Hong Kong | 3 |
| US DOJ Announces Four Cryptocurrency-Related Fraud Prosecutions | 3 |
| A Quick Glance At The Role Of The New EU-wide Authority On AML-CFT: The Anti-Money Laundering Authority | 3 |
| SIM registration to boost delivery of financial services devoid of fraud, theft – 1st Dep Governor | 3 |
| Federal government using artificial intelligence to fight cybercrime | 3 |
| Gambling giant Entain could lose UK licence after record £17m fine | 3 |
| Cost of investigations and increased fraud risk affect business stability | 3 |
| EDITORIAL / Financial loopholes must be closed to clean up money laundering | 3 |
| Japan’s police to increasingly partner up with AI to fight crime | 3 |
| TEAMING UP TO FIGHT FINANCIAL CRIME | 3 |
| Payment Monitoring Market To Grow Extensively at Unstoppable Rate \| ACI Worldwide, BAE Systems, Fiserv | 3 |
| UK Economic Crime Group: Enforcement Update - April 29, 2021 | 3 |
| 6AMLD Deadline is Nearing: A Look Back on Key Changes in Legal Framework Since 1AMLD | 3 |
| Worldwide Anti-money Laundering Solutions Industry to 2026 - Increased Adoption of Digital/Mobile Payment Solutions is Driving Growth... | 3 |
| Money laundering and jurisprudence | 3 |
| ‘India must raise the bar on anti-money laundering systems’ | 3 |
| Deference to those with influence fuels money laundering | 3 |
| U.K. Banks Adopt Digital Platforms to Compete with FinTechs | 3 |
| Nasdaq to Acquire Antifraud Company | 3 |
| Morning Digest \| Supreme Court orders immediate action against hate speech, Pakistan is out of FATF ‘grey list’ on terror funding, and more | 3 |
| Anti-Money Laundering Council; Analyser, Consultancy Service, Data Entry, Diversified Computer Hardware And Peripherals, Man Power, Other Software | 3 |
| JP Morgan partners AI/ML platform Cleareye to digitise trade finance operations | 3 |
| Insight EU monitoring - Your Daily Premium Digest | 3 |
| FinTech Global FS Regulatory Round-up – w/e 24 June 2022 | 3 |
| Machine Learning can make financial fraud prediction and detection simpler | 3 |
| House Financial Services: Task Force on Financial Technology: Whats in Your Digital Wallet? A Review of Recent Trends in Mobile Banking and... | 3 |
| Fintech – Ireland: Law And Practice | 3 |
| Regulators warn of new styles of high-tech fraud | 3 |
| Fintech trends: AI top term on Twitter February 2022 | 3 |
| This is what banking will look like in 10 years; From AI to crypto to sustainability | 3 |
| What Future AML Compliance Requires: A Technology Perspective | 3 |
| Lloyd's Selects 11 Start-Ups for Innovation Lab Accelerator | 3 |
| Luddite scare stories cannot be allowed to prevent British tech flourishing | 3 |
| Bank money laundering efforts increasingly impact on ordinary account holders | 4 |
| DNB's instructions to bunq partially deleted: no evidence for all anti-money laundering violations | 4 |
| From Indonesia to India: Is There Hope for Anti-Corruption Efforts Within the G20? | 4 |
| Factors To Be Considered By Financial Organizations For An Effective Transactional Monitoring System | 4 |
| South Africa banks face growing risks as anti-money laundering deadline looms | 4 |
| Seprelad detects "high importance" alerts in sports clubs | 4 |
| From Brazil to Dubai, Israeli startup secures billions of dollars from money laundering | 4 |
| AI And Algorithms In Financial Services—Future Areas Of Focus | 4 |
| BaFin To Focus On Six Main Risks And Two Future Risks In 2022 | 4 |
| A Quick Glance At The Role Of The New EU-wide Authority On AML-CFT: The Anti-Money Laundering Authority | 4 |
| Perkins Coie LLP - US Treasury Highlights Anti-Money Laundering Priorities in 2022 Illicit Finance Strategy | 4 |
| Know Future Outlook of Anti-money Laundering Software Market: Key Players EastNets, Fiserv, Opentext, Experian | 4 |
| Dubai Police monitor cryptocurrencies for anti-money laundering and crimes | 4 |
| Close coordination and supervision of financial intelligence units helped Macau reach highest tier of anti-money laundering jurisdictions... | 4 |
| The Contact Center's AI Revolution: AI technology is driving a rapid pace of innovation and change in the service sector.(REALITY CHECK) | 4 |
| A double-edged sword – Anti-money laundering regulations in the Pacific | 4 |
| Bank warns of hike in 'money mule' cases | 4 |
| Innovation, Communication Needed to Address Emerging Illicit Finance Threats, Say Regulators; Shifting regulatory obligations and new... | 4 |
| Feedzai's market leadership endorsed with category recognition by leading analyst firms | 4 |
| Are we friend or foe in the fight against financial crime? | 4 |
| MIL-OSI Banking: Protect The Poor When Fighting Money Laundering in the Pacific — by Lotte Schou-Zibell | 4 |
| Hong Kong's dirty money sleuths track criminals using cryptocurrency trading platforms to move illegal funds across borders | 4 |
| Bci Miami Re-Signs With QuantaVerse, Extends Use of Industry-Leading Financial Crime Investigation Platform | 4 |
| Anti Money Laundering Market is ready for its next Big Move \| Fiserv, Inc, Oracle Corporation, BAE Systems, Accenture | 4 |
| UK Economic Crime Group: Enforcement Update - April 29, 2021 | 4 |
| TO WHAT EXTENT CAN BLOCKCHAIN HELP DEVELOPMENT CO-OPERATION ACTORS MEET THE 2030 AGENDA? No. of pages: 50 | 4 |
| Review, Then Reform? AMLA Charts A Path For The Future Of SARs And CTRs | 4 |
| 6AMLD Deadline is Nearing: A Look Back on Key Changes in Legal Framework Since 1AMLD | 4 |
| NLP for Regulatory Compliance: A Question of When, Not If | 4 |
| DNB's instructions to bunq partially deleted: no evidence for all anti-money laundering violations | 4 |
| Global Anti-Money Laundering Solutions Market - Growth, Trends, COVID-19 Impact, and Forecasts (2022 - 2027) | 4 |
| HSBC Expands Partnership with Silent Eight to Fight Financial Crime | 4 |
| Santiago Nieto Castillo: New technologies and financial inclusion | 4 |
| Correspondent banking, a pending issue for SHCP and Banxico | 4 |
| From Indonesia to India: Is There Hope for Anti-Corruption Efforts Within the G20? | 4 |
| Factors To Be Considered By Financial Organizations For An Effective Transactional Monitoring System | 4 |
| South Africa banks face growing risks as anti-money laundering deadline looms | 4 |
| From Brazil to Dubai, Israeli startup secures billions of dollars from money laundering | 4 |
| Allied Market Research; Insurance Fraud Detection Market to Reach $28.1 Bn, Globally, by 2031 at 24.2% CAGR: Allied Market Research | 4 |
| Tax Authorities detect USD 1.5 million evasion in Tabesa's distribution companies | 4 |
| Mint Explainer: How India is lifting tax collections with a data drive | 4 |
| Video: How Payoneer prevents financial crime with ThetaRay’s tech | 4 |
| NetGuardians moves into money laundering transaction monitoring | 4 |
| SIM registration to boost delivery of financial services devoid of fraud, theft – 1st Dep Governor | 4 |
| Curbing money laundering through AI | 4 |
| Government must make sure regulation of fintech does not stifle innovation in Hong Kong | 4 |
| AML partners urge operationalisation of AI to solve SA’s pressing AML challenges | 4 |
| Impesa implements technologies to contribute to the fight against the entry of illicit money into the National Financial System | 4 |
| Hawk AI appoints FinCrime veteran Steve Liu as General Manager for the USA | 4 |
| Know Future Outlook of Anti-money Laundering Software Market: Key Players EastNets, Fiserv, Opentext, Experian | 4 |
| Financial institutions turn to AI and ML to aid anti-money laundering efforts | 4 |
| Fighting Financial Crime with Advanced Technologies | 4 |
| AMLC: Philippines to use AI to fight dirty money | 4 |
| Anti Money Laundering Market is ready for its next Big Move \| Fiserv, Inc, Oracle Corporation, BAE Systems, Accenture | 4 |
| Anti-money Laundering Software Market to see Huge Growth by 2025 \| Trulioo, Verafin, Fenergo | 4 |
| Federal Agencies Clarify Application Of Model Risk Management Guidance To BSA/AML Compliance And Request Information From Industry | 4 |
| How AML compliance is intersecting with payment fraud prevention | 4 |
| Microsoft eyeing Discord; Amazon’s expansion; tech against money laundering | 4 |
| Anti-money Laundering Software Market to See Huge Growth With Accenture Inc., SAS Institute Inc., Fiserv | 4 |
| Anti-money Laundering Market Size, Share & Trends Analysis Report By Component (Software, Services), By Product Type, By Deployment, By End... | 4 |
| Anti-Money Laundering Market Size Will Reach USD 2385.8 Million by 2026: Facts & Factors | 4 |
| Federal Banking Agencies Seek Greater Flexibility In Granting Exemptions From SAR Requirements | 4 |
| Money laundering reform from Congress is needed more than ever | 4 |
| Only 1% of laundered cash in EU is detected — ABN AMRO wants to improve that; ABN AMRO's Ivich Hoffman reckons detecting money laundering is... | 4 |
| 62% of financial institutions reported an increase in financial crimes | 4 |
| OFAC Issues New Sanctions Compliance Guidance For Instant Payment Systems | 4 |
| Artificial Intelligence Stocks To Watch: Big Tech Expands AI Products, Services | 4 |
| Artificial intelligence helps to reduce fraud and financial crimes | 4 |
| Banks Turn to AI to Help Dodge Enforcement Spotlight; Regulators increasingly expect banks and other entities to use smarter systems to help... | 4 |
| Bank regulators need to overhaul their digital capacity | 4 |
| Artificial Intelligence reduces non-payment risks | 4 |
| Global Industry Snapshot - GIS0001310 The Impact of Fintech on Financial Services 14 March 2022 | 4 |
| RBI fintech and banking sector predictions for 2022 | 4 |
| China’s central bank governor calls for personal data protection at Hong Kong FinTech Week | 4 |
| AI in fintech market to Witness Significant Incremental Opportunity By 2027- AllTheResearch | 4 |
| Anti Money Laundering Market is ready for its next Big Move \| Fiserv, Inc, Oracle Corporation, BAE Systems, Accenture | 4 |
| Anti-money Laundering Software Market to see Huge Growth by 2025 \| Trulioo, Verafin, Fenergo | 4 |
| Federal government using artificial intelligence to fight cybercrime | 4 |
| Global Anti-Money Laundering Solutions Market - Growth, Trends, COVID-19 Impact, and Forecasts (2022 - 2027) | 5 |
| ThetaRay Releases Advanced SONAR AML Solution with Upgrades for Fintechs and Banks | 5 |
| Why U.S. Law Firms Should Start Preparing For Money Laundering Regulation | 5 |
| Government must make sure regulation of fintech does not stifle innovation in Hong Kong | 5 |
| Amid pressure to fight money laundering, Treasury looks to spur AI innovation | 5 |
| ThetaRay; ThetaRay AI Technology to Monitor Knox Wire Cross-Border Payments | 5 |
| AML Compliance Strategies for Crypto Exchanges; COMPLIANCE | 5 |
| NICE Limited - NICE Actimize Honored for Best Anti-Money Laundering Solution by Regulation Asia’s Excellence Awards | 5 |
| Global Anti-Money Laundering Tools Market, By Component (Software, Services); By organization Size (Small and Medium Enterprises (SMEs... | 5 |
| Bank Negara Malaysia: Deputy Governor's Welcoming Remarks at the 23rd Asia/Pacific Group on Money Laundering (APG) Typologies Workshop | 5 |
| Excessive anti-money laundering regulation slows financial inclusion; little criminal prosecution | 5 |
| Not just NatWest: Compliance guru on why money laundering is rampant in the City | 5 |
| TEAMING UP TO FIGHT FINANCIAL CRIME | 5 |
| Financial institutions turn to AI and ML to aid anti-money laundering efforts | 5 |
| AMLC: Philippines to use AI to fight dirty money | 5 |
| Nedbank’s Tjipuka fighting illicit financial crimes | 5 |
| Anti-money Laundering Software Market to see Huge Growth by 2025 \| Trulioo, Verafin, Fenergo | 5 |
| MFS Africa Selects ThetaRay to Monitor Money Laundering and Sanctions List Violations | 5 |
| ThetaRay Releases Advanced SONAR AML Solution with Upgrades for Fintechs and Banks | 5 |
| Giant Oak Inc. Giant Oak Releases New Updates to GOST, Enhancing Risk Assessment to Identify Corruption, Terrorism, Money Laundering, and... | 5 |
| 5 Tools Data Scientists Need to Help Combat Financial Crimes | 5 |
| 8% of Latin American companies use Artificial Intelligence for human talent management | 5 |
| Financial Crime Is More Common And Innovative Than Ever: Here’s What To Know | 5 |
| Target Group chooses NICE Actimize for AML | 5 |
| Need for Better AML Compliance in the Netherlands | 5 |
| HSBC and Silent Eight Expand Machine Learning Partnership | 5 |
| FINOM Selects Resistant AI's Transaction Forensics to Strengthen its Money Laundering Defences | 5 |
| AI And Algorithms In Financial Services—Future Areas Of Focus | 5 |
| Banks can expect return to strict oversight of anti-money laundering rules | 5 |
| Virtusa Corporation - Virtusa and Major US University’s Collaboration Develops AI Technologies that Address Fraud, Money Laundering a... | 5 |
| IRS turns to data analytics to track crypto tax evasion | 5 |
| KPMG: Nigerian Banks Spend Billions Annually on KYCs | 5 |
| Fraud, Asset Tracing & Recovery 2021/2022 – Guernsey (Commercial Dispute Resolution, CDR) | 5 |
| Fintech startup raises $200 million to fight digital banking fraud | 5 |
| QuantaVerse Launches New Financial Crime Investigation Report to Streamline Level 1 AML Investigations | 5 |
| UOB’s new AI anti-money laundering solution helps the Bank cut through large volumes of transactions to pinpoint suspicious activities | 5 |
| Saudi Arabia has in place ‘measures to combat terrorist financing’ | 5 |
| Bank of Greece - Speech by the Bank of Greece Governor Yannis Stournaras at the 7th Annual Conference of Mediterranean Central Banks | 5 |
| RCBC and GBG win two awards for AI and fraud prevention initiatives at the Asian Banking & Finance Retail Banking Awards 2022 | 5 |
| Deutsche Bundesbank - Machine learning in risk models – brontobytes and Brontosauruses Speech delivered at the Bundesbank-BaFin FinTech Festival | 5 |
| Global Industry Snapshot - GIS0001426 The Impact of Fintech on Financial Services 16 May 2022 | 5 |
| Fintech trends: Artificial intelligence top mentioned term on Twitter Q4 2021 | 5 |
| Anticipating the road ahead | 5 |
| Swedish Finance Group Establishes New Anti-Money Laundering Compliance Unit | 5 |
| Fintech in the UAE | 5 |
| Deltec Bank, Bahamas Says Artificial Intelligence Is Transforming Fraud Prevention In Banking and Finance Sector | 5 |

Analysis of these ratings produced the following distributions:

| Sentiment Score | Count |
|---|---|
| 1 | 5 |
| 2 | 17 |
| 3 | 57 |
| 4 | 78 |
| 5 | 46 |
| Grand Total | 203 |

| Calculation | Value |
|---|---|
| Average | 3.7 |
| Median | 4 |

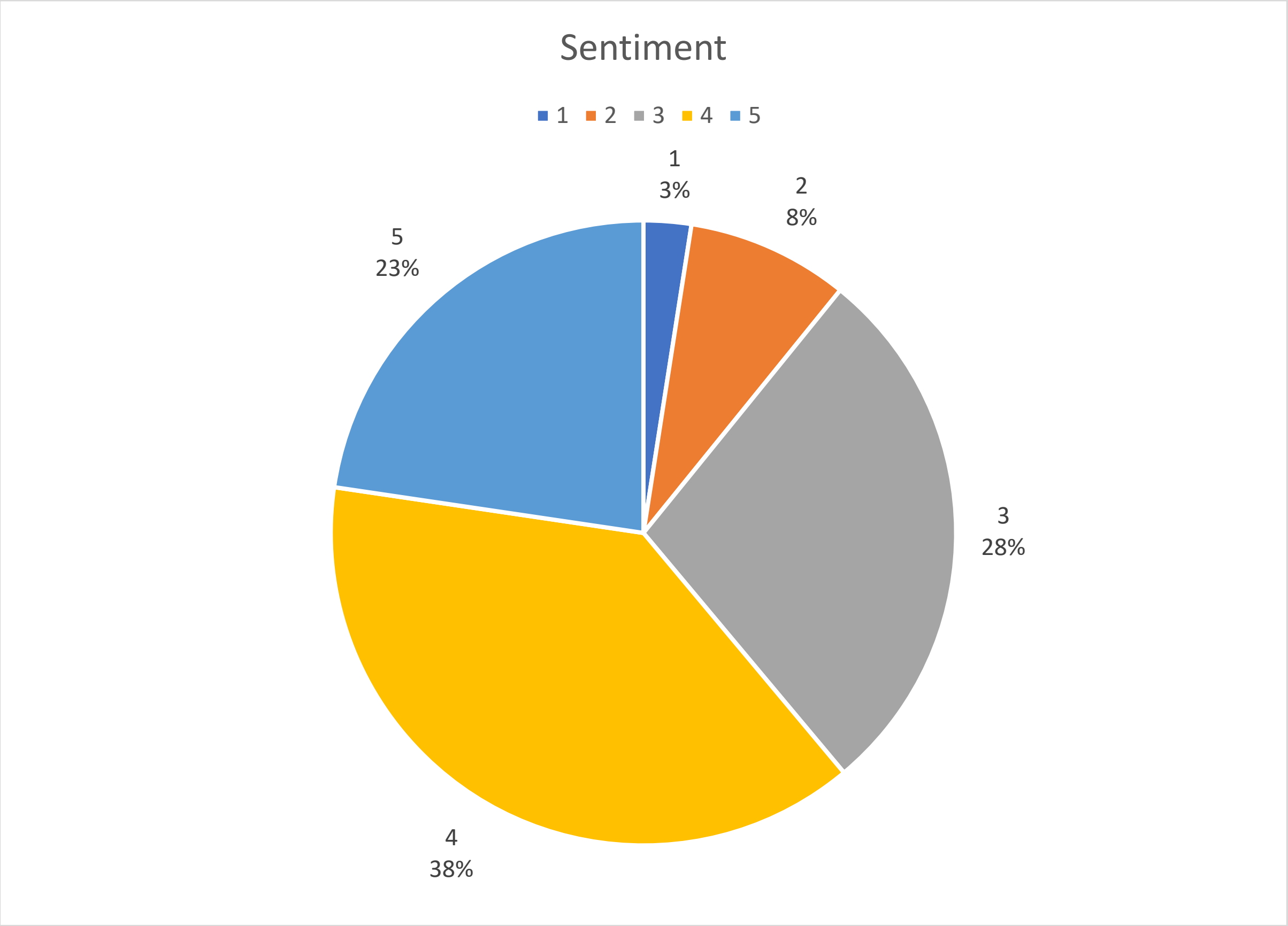

As can be seen from this chart, the positive sentiment in favour of using artificial intelligence to fight money laundering makes up 61% of the results. Moreover, this large proportion of highly positive articles (sentiment score 5) does not mention any downside or drawbacks to the technology.

Further sentiment evidence is clear from the titles of the articles themselves. A 30-word word cloud derived from the article titles shows the importance of specific keywords used in the news reporting:

*Figure 10 - Word cloud of sentiment results titles*

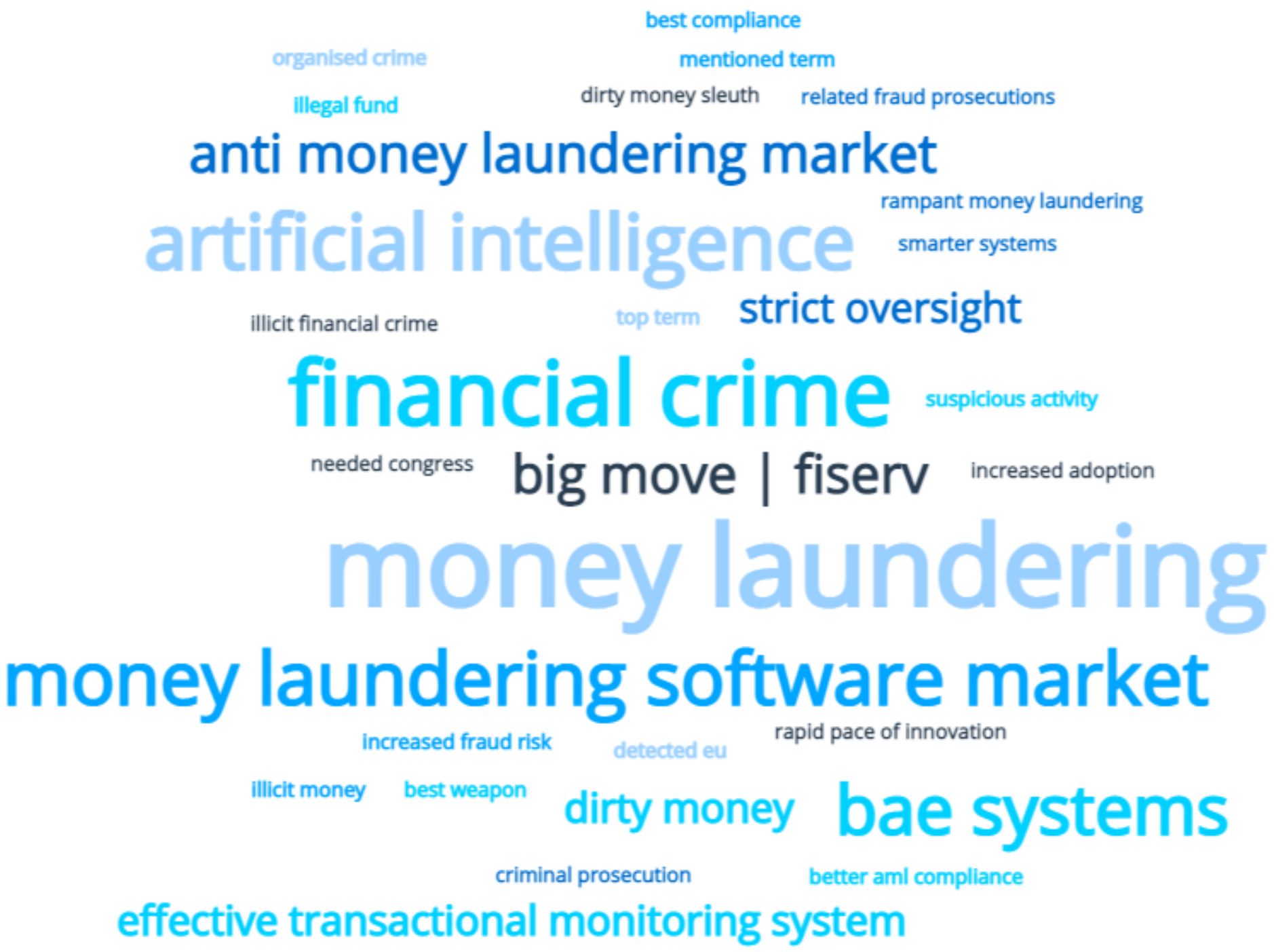

Note that this word cloud contains little in terms of warnings re the use of artificial intelligence solutions in anti-money laundering.

**Questionnaire findings**

This method is to go directly to those labouring under a regulatory burden for their impressions and opinions.

The findings from the questionnaire are as follows:

| Number of Respondents |
|---|
| 104 |

**Section A – Demographics**

## *Question 1*

Q1. Which of the following best describes your current occupation?

| Answer Choices | Response Percent | Responses |
|---|---|---|
| Financial crime prevention (Anti-Money Laundering) | 21.15% | 22 |
| MLRO or CCO | 24.04% | 25 |
| Consultant (Fin Crime) | 12.50% | 13 |
| Consultant (Other) | 6.73% | 7 |
| Technology vendor (FinTech) | 7.69% | 8 |
| Technology vendor (Other) | 12.50% | 13 |
| Bank employee | 3.85% | 4 |
| Data Scientist or Machine Learning engineer | 1.92% | 2 |
| Legal, regulatory or law enforcement | 0.96% | 1 |
| Sales executive | 0.96% | 1 |
| Other (please specify) | 7.69% | 8 |
| | Answered | 104 |
| | Skipped | 0 |

| Other (please specify) |
|---|
| DMLRO |
| DMLRO |
| Compliance Examiner |
| Data Analyst |
| KYC Analyst at a bank |
| Researcher |
| Politically Exposed Persons Specialist |
| Risk Manager |

Note that the top two results are people working in AML and MLROs. A DMLRO works as a deputy to the MLRO role. Many of these roles directly work with compliance data (except "sales" roles) and therefore represent the core audience for the research.

### *Question 2*

Q2. Which of the following best describes your current job level?

| Answer Choices | Response Percent | Responses |
|---|---|---|
| Owner/Executive/C-Level | 12.50% | 13 |
| Senior Management | 41.35% | 43 |
| Middle Management | 31.73% | 33 |
| Intermediate | 9.62% | 10 |
| Entry Level | 3.85% | 4 |
| Other | 0.96% | 1 |
| | Answered | 104 |
| | Skipped | 0 |

Note that most respondents are senior employees, demonstrating the emphasis on the MLRO role being a senior one.

Combining these two results against each other shows that, where the answer to question 2 was "Senior Management", the most common answer to question 1 was "MLRO or CCO":

*Table 1 - Q1 + Q2, where Q1 "Senior Management"*

| Which of the following best describes your current job level? | Senior Management |
|---|---|
| **Role** | **Count** |
| MLRO or CCO | 17 |
| Financial crime prevention (Anti-Money Laundering) | 6 |
| Technology vendor (Other) | 5 |
| Consultant (Fin Crime) | 5 |
| Consultant (Other) | 4 |
| Technology vendor (FinTech) | 3 |
| Bank employee | 1 |
| Data Scientist or Machine Learning engineer | 1 |
| **Grand Total** | **42** |

Furthermore, where the answer was "Intermediate", "Middle Manager", or "Other", the answer was "Financial crime prevention (Anti-Money Laundering)":

*Table 2 - Q1 + Q2, where Q1 “Intermediate”, “Middle Manager”, or “Other”*

| Which of the following best describes your current job level? | (Multiple Items) |
|---|---|

| **Role** | **Count** |
|---|---|
| Financial crime prevention (Anti-Money Laundering) | 14 |
| Technology vendor (Other) | 7 |
| Other (please specify) | 6 |
| Technology vendor (FinTech) | 4 |
| Consultant (Fin Crime) | 3 |
| Bank employee | 3 |
| MLRO or CCO | 2 |
| Consultant (Other) | 2 |
| Data Scientist or Machine Learning engineer | 1 |
| Sales executive | 1 |
| Legal, regulatory or law enforcement | 1 |
| **Grand Total** | **44** |

### *Question 3*

Q3. What is your age?

| Answer Choices | Response Percent | Responses |
|---|---|---|
| Under 18 | 0.00% | 0 |
| 18-24 | 0.96% | 1 |
| 25-34 | 25.96% | 27 |
| 35-44 | 43.27% | 45 |
| 45-54 | 17.31% | 18 |
| 55-64 | 11.54% | 12 |
| 65+ | 0.96% | 1 |
| | Answered | 104 |
| | Skipped | 0 |

Age was chosen explicitly as a grouping factor as it correlates with seniority. The sharply delineated age profile shown here is expected as working in the AML industry often requires qualifications not available at universities coupled with earned experience.

### *Question 4*

Q4. In what country do you work?

| Answer Choices | Response Percent | Responses |
|---|---|---|
| United Kingdom of Great Britain and Northern Ireland | 45.19% | 47 |
| United Arab Emirates | 8.65% | 9 |
| United States of America | 8.65% | 9 |
| Singapore | 6.73% | 7 |
| Netherlands | 5.77% | 6 |
| Australia | 2.88% | 3 |
| Spain | 2.88% | 3 |
| Portugal | 1.92% | 2 |
| Qatar | 1.92% | 2 |
| Saudi Arabia | 1.92% | 2 |
| Afghanistan | 0.96% | 1 |
| Austria | 0.96% | 1 |
| Canada | 0.96% | 1 |
| Denmark | 0.96% | 1 |
| Egypt | 0.96% | 1 |
| Germany | 0.96% | 1 |
| India | 0.96% | 1 |
| Ireland | 0.96% | 1 |
| Italy | 0.96% | 1 |
| Jamaica | 0.96% | 1 |
| Kuwait | 0.96% | 1 |
| Lebanon | 0.96% | 1 |
| New Zealand | 0.96% | 1 |
| Sweden | 0.96% | 1 |
| | Answered | 104 |

**Section B – Regulations**

## *Question 5*

Q5. "Regulations significantly burden companies and those involved in combatting financial crime."

| Answer Choices | Response Percent | Responses |
|---|---|---|
| Strongly agree | 16.90% | 12 |
| Agree | 30.99% | 22 |
| Neither agree nor disagree | 15.49% | 11 |
| Disagree | 26.76% | 19 |
| Strongly disagree | 9.86% | 7 |
| | Answered | 71 |
| | Skipped | 33 |

This question was deliberately provocative to challenge the respondent to consider their feelings. The author received one complaint about the language use of “burden” as “very negative”. However, the language of question 5, specifically "burden", is taken directly from the UK government in their "Action Plan for anti-money laundering and counter-terrorist finance".

Question 5 was a key question for correlations. “Strongly agree” accounted for 16.9% of responses and when this value was correlated with Q6, then “The burden has increased” was the most common answer (83.33%):

*Table 3 - Q5 "Strongly agree" + Q6*

Q6 Do you think the regulatory burden has changed significantly in the last five years?

Answered: 12 Skipped: 0

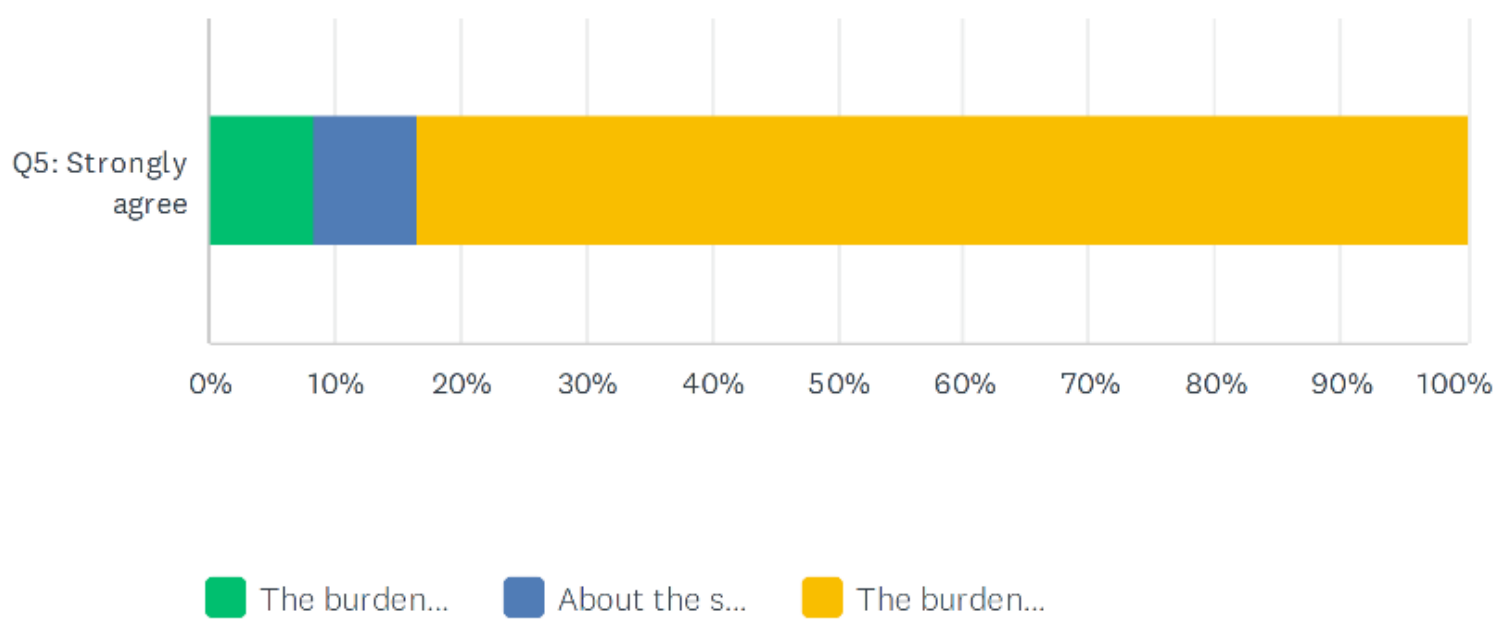

| | THE BURDEN HAS REDUCED | ABOUT THE SAME | THE BURDEN HAS INCREASED | TOTAL |
|---|---|---|---|---|
| Q5: Strongly agree (A) | 8.33%<br>1 | 8.33%<br>1 | 83.33%<br>10 | 100.00%<br>12 |
| Total Respondents | 1 | 1 | 10 | 12 |

This provides evidence for Hypothesis 2: “Questionnaire responses show significant evidence of the impacts of regulation placing an increasing burden on the regulated.”

Similarly, when paired with Question 9:

*Table 4 - Q5 "Strongly agree" + Q9*

Q9 "The volume or rate of Suspicious Activity Reports has increased significantly since I entered the business."

Answered: 12 Skipped: 0

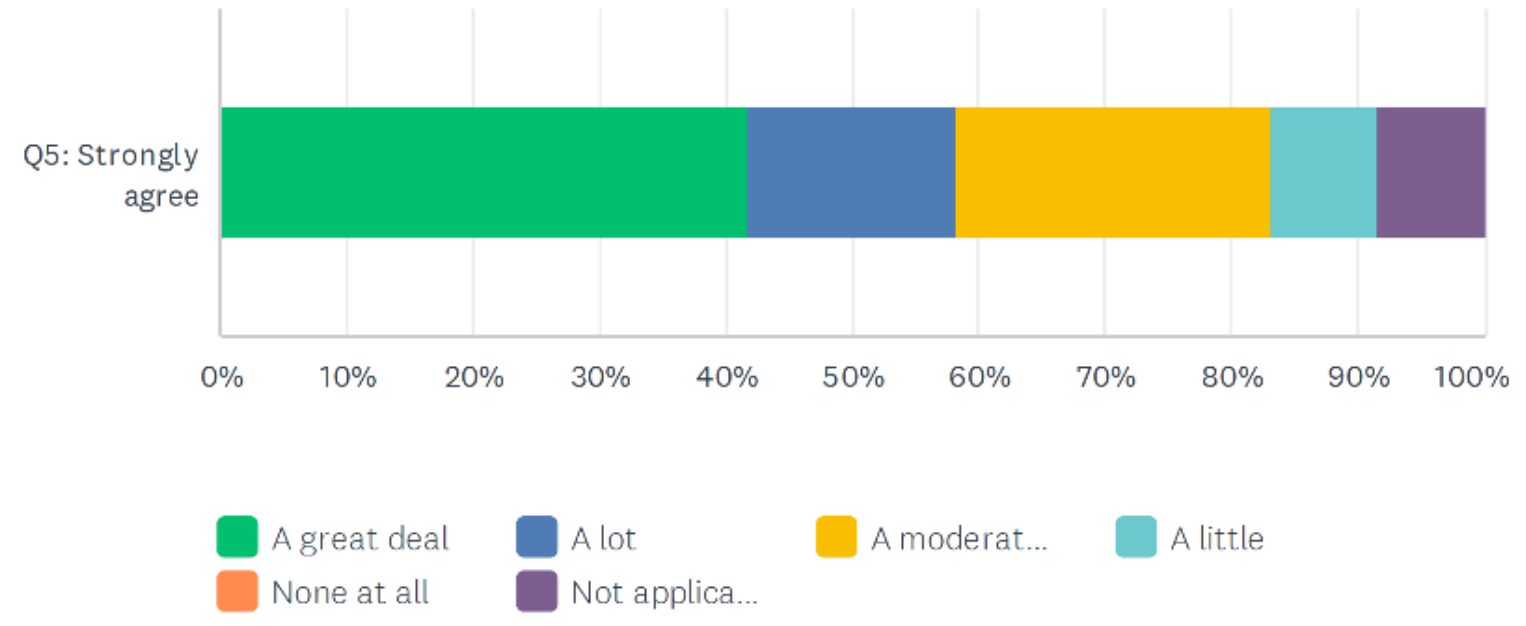

| | A GREAT DEAL | A LOT | A MODERATE AMOUNT | A LITTLE | NONE AT ALL | NOT APPLICABLE | TOTAL |
|---|---|---|---|---|---|---|---|
| Q5: Strongly agree (A) | 41.67%<br>5 | 16.67%<br>2 | 25.00%<br>3 | 8.33%<br>1 | 0.00%<br>0 | 8.33%<br>1 | 100.00%<br>12 |
| Total Respondents | 5 | 2 | 3 | 1 | 0 | 1 | 12 |

Furthermore, this group also found it took significant time to manage the matches produced in AML systems:

*Table 5 - Q5 "Strongly agree" + Q10*

Q10 The amount of review matches in AML systems requires significant time to manage.

Answered: 12 Skipped: 0

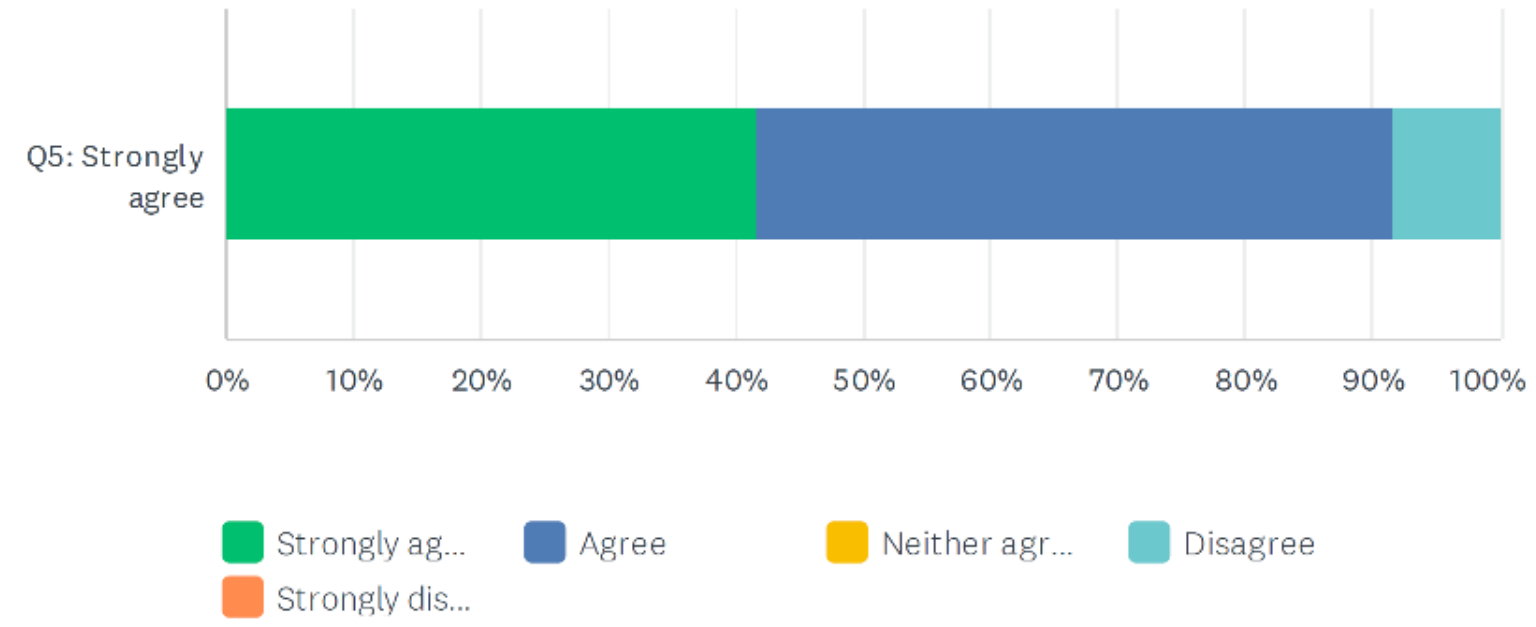

| | STRONGLY AGREE | AGREE | NEITHER AGREE NOR DISAGREE | DISAGREE | STRONGLY DISAGREE | TOTAL |
|---|---|---|---|---|---|---|
| Q5: Strongly agree (A) | 41.67%<br>5 | 50.00%<br>6 | 0.00%<br>0 | 8.33%<br>1 | 0.00%<br>0 | 100.00%<br>12 |
| Total Respondents | 5 | 6 | 0 | 1 | 0 | 12 |

Finally, the largest Question 5 group or "Agree", were also the largest group to mention the use of outsourcing to offshore (Low-Income Countries) for match reviews (Question 12):

*Table 6 - Q5 "Agree" + Q12*

| "Regulations significantly burden companies and those involved in combatting financial crime." | Agree |
|---|---|

| **"Have you required the use of outsourced or offshore staff to handle volumes of matches?"** | **Count** |
|---|---|
| Yes | 11 |
| Other (please specify) | 6 |
| No | 5 |
| **Grand Total** | **22** |

This suggests that the use of large numbers of offshore labourers to deal with the increase in the regulatory burden is a typical response to the fundamental problems thrown up in the literature review chapter.

Reviewing Question 6 highlights that only 5.63% of respondents believed the burden to have reduced at all:

### *Question 6*

Q6. Do you think the regulatory burden has changed significantly in the last five years?

| Answer Choices | Response Percent | Responses |
|---|---|---|
| The burden has reduced | 5.63% | 4 |
| About the same | 22.54% | 16 |
| The burden has increased | 71.83% | 51 |
| | Answered | 71 |
| | Skipped | 33 |

Question 7 allowed the respondents to input why they thought that this change was in a free form open question format:

### *Question 7*

Q7. If so, why?

| | |
|---|---|
| Answered | 61.00 |
| Skipped | 43.00 |

| Responses |
| --- |
| Australian regulators finally started caring |
| Cause I have seen same staff over years. |
| Requirements have been constant, but activity has increased - it's a scale issue e.g. sanctions compliance |
| It is not the rules, but poor supervision |
| Ongoing new regulations |
| Reforms |
| Business and operations has changed and compliance became Business as usual |
| Regulators competing, regulatory brain drain, new products and services eg crypto being developed for public use |
| More requested by law |
| BO discrepancy reporting. Greater emphasis on Business Risk Assessment. |
| Enhanced regulations |
| Whilst the risk based approach was introduced, rules have become more prescriptive (register of overseas entities, discrepancy reporting to companies House etc) |
| Number of Regulations, duties and requirements increased |
| More awareness and regulators are more serious |
| Ever since Saudi was trying to be a permanent FATF member, local AML regulations within the financial industry faced lots of positive changes, as rules growing and widen, technology and infrastructure within the country made it easier and secure more than any time before |
| Regulatory requirements need to be applied across all institutions regardless of size and scope of products and services |
| Increase in sanctions, increase in cross boarder regulators sharing |
| Regulators applying pressure and increasing inspections |
| More regulations have come in, more "box ticking" to be done by financial institutions |
| Crypto assets have generated an uncharted territory |
| I think regulations take longer to interpret and implement. They are also open to interpretation in many cases |
| Current world events and increasing focus on sanctions |
| Events that have resulted in significant financial gains by criminals. |
| No significantly new regulatory principles and requirements, but significant changes in complexity of external environment (sanctions, payment technologies, crypto currencies) |
| n/a, about the same but there should be more |
| More regulation is in placed, more checks required to be done |
| The regulatory tightening the requirements due to regulatory failures, and pressure from government. |
| As regulatory bodies implement new processes, firms are then required to implement these regulatory requirements which cause an impact on operations |
| In the Netherlands banks are now allowed to exchange client information to combat financial crime. This is a huge step due to privacy laws. |
| new legal acts and regulations put in place |
| Advancement in technology has provided new opportunities to criminals to launder money. The global digitalisation is leading to more cyber attacks and ransomware attacks. The introduction of cryptocurrencies is adding more risk as they are decentralised and unregulated. The unfolding events in Russia are also placing regulatory authorities in leading financial centres under increasing pressure to crack down on money laundering linked to the region. Lastly |
| Feels about the same, Banks are governance heavy anyway, no change in perception |
| Reaction to public reports and private experiences related to being the target of and/or unwittingly involved in financial crimes such as money laundering. |
| The number of sanctioned people has significantly increased in the last year. Plus, more and more countries are developing AML regulations and they can differ from one country to another, |
| so companie operating internationally need to adapt to all these particularities and changes. They also constantly need to screen more and more people and companies. |
| The regulators are taking a much tougher line and are quicker to take action |
| Overall expectations have increased of regulators |
| More proactive detection with technology |
| There are additional regulations and larger fines. |
| Nothing has really changed from a regulatory POV, firms have made it more complicated. |
| scrutiny is still high, haven't seen a marked increase, just more diligence in the regulator's understanding of more growing areas of fincrime risk |
| Same principles, more prescriptive processes |
| More regulation |
| The regulations in the middle east over the past 5 years have gone from basic to more elaborate, which has created a larger burden than before (although this now lines up with US/EU) |
| more regs; stronger enforcement |
| Financial Crime is not limited to paper currency only |
| FATF |
| Regulators getting slightly smarter and enforcing more, meaning companies need to do more to comply with rules that are essentially pretty similar to how they have been for years. |
| Companies are not in the business of crime detection or prevention. Companies does not have the capacity to be an enforcement unit. Companies can assist however the burden has been increased to such extent that the onus lies with the private sectors. |
| The rise of crypto currencies, financial inclusion, open banking |
| The regulatory environment is always changing |
| Not a burden but a support as it simplifies the frameworks to apply |
| Lots of data, records and client info is requested by the regulators. |
| Regulators feel under pressure to stamp out poor behaviour |
| Public's perception |
| Criminals are exploiting the Banking system to faciliate illicit activity at massive scale. The cost to business and society has led to calls to tighten regulation and penalties. This is gradually occuring with the commensurate increase in resource required by industry. |
| Probably because AML results based on previous regulations was not satisfactory |
| Because the fight against financial crime and its enforcement were perceived as mainly for the financial sector while it is for financial & non-financial. |
| New regulations like checking UBO |
| more rules |
| There are some new regulatory requirements which are introduced recently; however, the advance in the technology and the availability of new tools means the burden has probably remained the same. |
| Divergence of sanctions into an ever-increasing complex set of lists and regimes. Needs to identify beneficial owners without readily available access to centralized registers globally and the lack of detailed guidance on gauging the reliability and political bias associated with different news outlets for using in Adverse Media screening. |

A word cloud drawn from these answers, highlights the keys point that regulations are increasing, further providing evidence for Hypothesis 2:

*Figure 11 - Word cloud of Q7 responses*

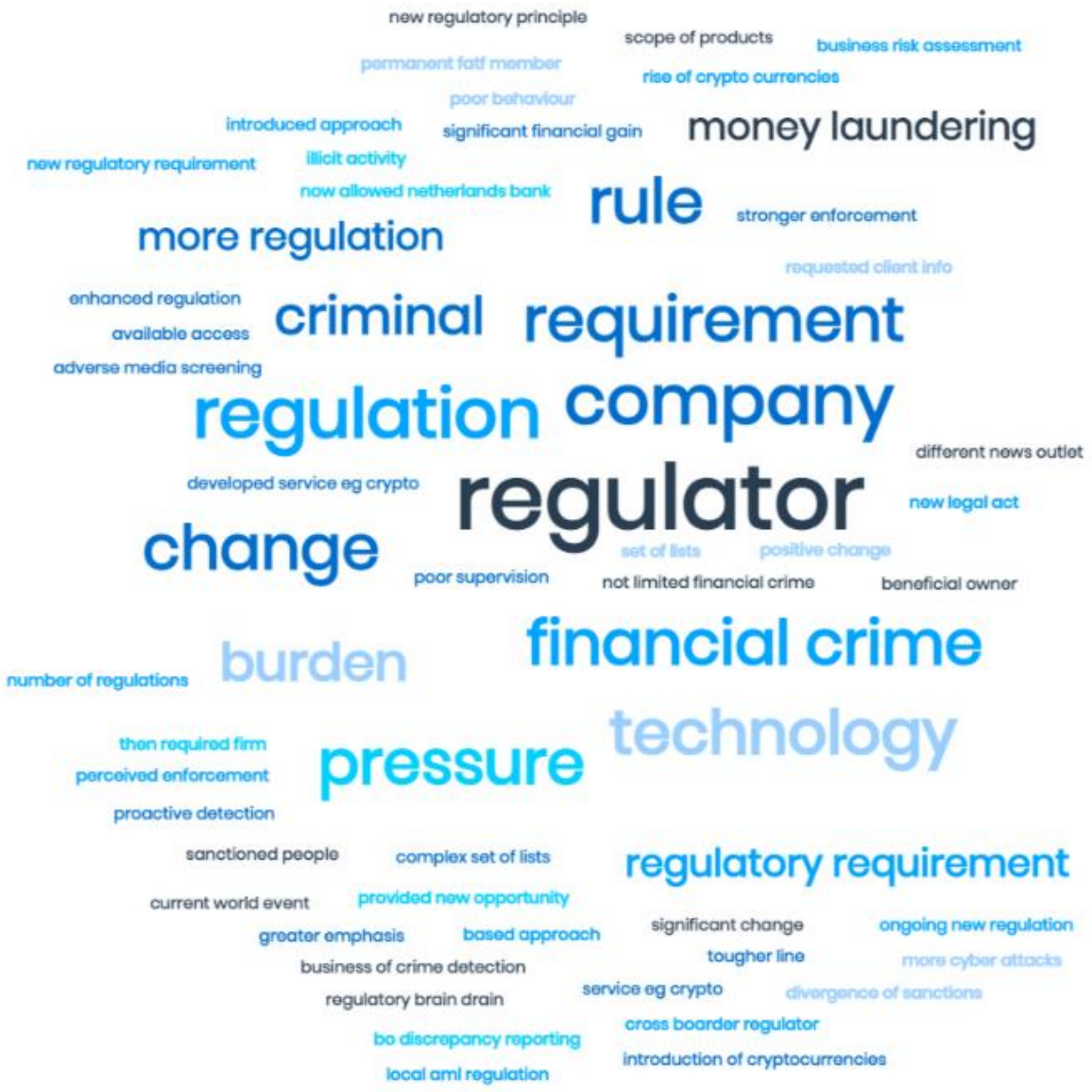

### *Question 8*

Q8. "AML methodologies have significantly changed for many years."

| Answer Choices | Response Percent | Responses |
|---|---|---|
| Strongly agree | 18.57% | 13 |
| Agree | 54.29% | 38 |
| Neither agree nor disagree | 12.86% | 9 |
| Disagree | 11.43% | 8 |
| Strongly disagree | 2.86% | 2 |
| | Answered | 70 |
| | Skipped | 34 |

Question 8 is asking if the paradigm has changed in the opinion of the respondent. The evidence from the literature review suggested that AML methods were stuck in a rut, to which AI is seen as the solution. However, the largest answer to this question is that they have indeed changed.

Correlation with demographics showed that this score was accounted for mainly by a single group, those working in Singapore:

*Table 7 - Q8 + Q4 "Singapore"*

| In what country do you work? | Singapore |
|---|---|
| **AML methodologies have significantly changed for many years.** | **Count** |
| Strongly agree | 4 |
| (blank) | 2 |
| Agree | 1 |
| **Grand Total** | **7** |

This group of respondents self-describe their occupation as:

*Table 8 - Q8 + Q4 "Singapore" + Q1*

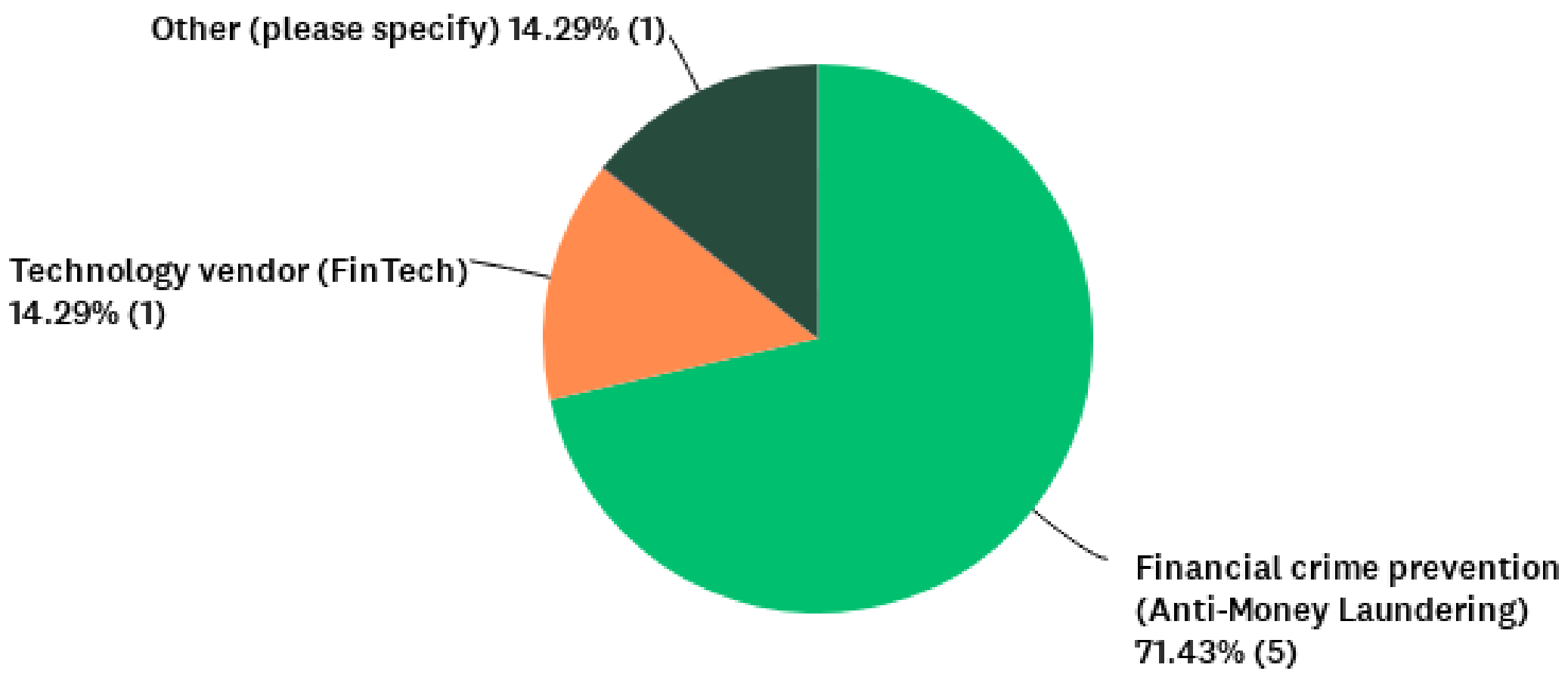

When combined with question 11, it demonstrates the proportion of the time this group spends on reviewing matches:

*Table 9 - Q8 + Q4 "Singapore" + Q11 (shown as % of total time working)*

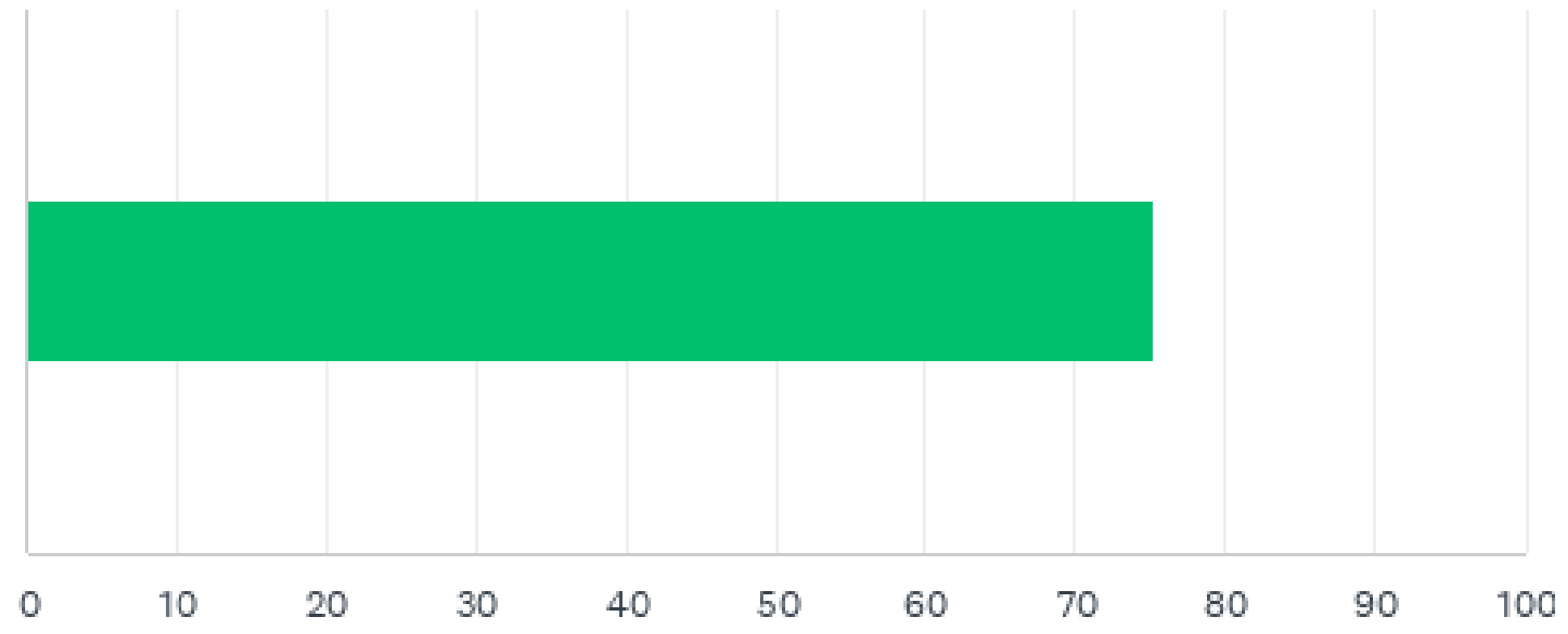

| ANSWER CHOICES | AVERAGE NUMBER | TOTAL NUMBER | RESPONSES |
|---|---|---|---|
| Responses | 75 | 376 | 5 |
| **Total Respondents: 5** | | | |

This very high result suggests that this group would be people working on matches within AML systems, either as outsourced workers or internal groups offshore.

### *Question 9*

Q9. "The volume or rate of Suspicious Activity Reports has increased significantly since I entered the business."

| Answer Choices | Response Percent | Responses |
|---|---|---|
| A great deal | 18.57% | 13 |
| A lot | 37.14% | 26 |
| A moderate amount | 21.43% | 15 |
| A little | 8.57% | 6 |
| None at all | 2.86% | 2 |
| Not applicable | 11.43% | 8 |
| | Answered | 70 |
| | Skipped | 34 |

The answers to this question provide evidence for the issues discovered in the literature review. A very high proportion of respondents think SARs are increasing, bearing out the information from the UKFIU.

### *Question 10*

Q10. The amount of review matches in AML systems requires significant time to manage.

| Answer Choices | Response Percent | Responses |
|---|---|---|
| Strongly agree | 31.43% | 22 |
| Agree | 51.43% | 36 |
| Neither agree nor disagree | 8.57% | 6 |
| Disagree | 8.57% | 6 |
| Strongly disagree | 0.00% | 0 |
| | Answered | 70 |
| | Skipped | 34 |

### *Question 11*

Q11. If you are involved with reviewing matches: what proportion of time is spent reviewing matches in your organisation or organisations you have dealt with?

| Answer Choices | Average Number | Total Number | Response Percent | Responses |
|---|---|---|---|---|
| (no label) | 35.12 | 1791 | 100.00% | 51 |
| | | | Answered | 51 |
| | | | Skipped | 53 |

### *Question 12*

Q12. Have you required the use of outsourced or offshore staff to handle volumes of matches?

| Answer Choices | Response Percent | Responses |
|---|---|---|
| Yes | 39.13% | 27 |
| No | 42.03% | 29 |
| Other (please specify) | 18.84% | 13 |
| | Answered | 69 |
| | Skipped | 35 |

### *Question 13*

Q13. "I think a more balanced approach to regulation and AML would be more productive to combatting financial crime."

| Answer Choices | Response Percent | Responses |
|---|---|---|
| Strongly agree | 13.04% | 9 |
| Agree | 43.48% | 30 |
| Neither agree nor disagree | 26.09% | 18 |
| Disagree | 15.94% | 11 |
| Strongly disagree | 1.45% | 1 |
| Other (please specify) | 0.00% | 0 |
| | Answered | 69 |
| | Skipped | 35 |

Question 13 is crucial to the likelihood of working for change in AML screening. Only one respondent felt more regulation was the answer to the current burden and volumes of SARs.

### *Question 14*

14. The FCA said in 2019 that money laundering causes "incalculable" damage to society and that technology might be the answer. Do you agree?

| Answer Choices | Response Percent | Responses |
|---|---|---|
| Strongly agree | 42.25% | 30 |
| Agree | 42.25% | 30 |
| Neither agree nor disagree | 14.08% | 10 |
| Disagree | 1.41% | 1 |
| Strongly disagree | 0.00% | 0 |
| | Answered | 71 |
| | Skipped | 33 |

All groups of respondents well recognise the dangers of money laundering.

### *Question 15*

Q15. Do you think the skills required to perform AML successfully are being lost due to the high burdens of the regulations?

| Answer Choices | Response Percent | Responses |
|---|---|---|
| A great deal of skills are being lost | 7.25% | 5 |
| A lot of skills are being lost | 13.04% | 9 |
| A moderate amount of skills are being lost | 27.54% | 19 |
| A little amount of skills are being lost | 15.94% | 11 |
| No skills are being lost due to regulation | 30.43% | 21 |
| Other (please specify) | 5.80% | 4 |
| | Answered | 69 |
| | Skipped | 35 |
| | | |

| Other (please specify) |
|---|
| It's not that skills are lost. It's that time is lost because you can't access data or documents. Additionally banks have to do an independent research of the same client which costs time and money. |
| Not really. The level of scrutiny may put off some people who have responsibility but generally the high level of enforcement and focus means that there are lots of jobs, which means much training and development over the last 10-20 years. |
| 1st line employees fall short in recognizing AML risks. That is partly a lack of skill, partly a lack of motivation. |
| also, because old/silos legacy technology & processes. |

**Section C – AI in AML**

This section of the questionnaire focussed on using AI as a technology in AML, providing evidence mainly for Hypotheses 1 and 3.

### *Question 16*

Q16. "I would say that AML technology has kept pace with regulation changes."

| Answer Choices | Response Percent | Responses |
|---|---|---|
| Strongly agree | 2.86% | 2 |
| Agree | 47.14% | 33 |
| Neither agree nor disagree | 20.00% | 14 |
| Disagree | 28.57% | 20 |
| Strongly disagree | 1.43% | 1 |
| | Answered | 70 |
| | Skipped | 34 |

This question measured the perception of AML technology concerning the regulation changes. Could there be a correlation between those who think the regulatory burden is great?

*Table 10 - Q5 "Strongly agree" + Q16*

| "Regulations significantly burden companies and those involved in combatting financial crime." | Strongly agree |
|---|---|
| **I would say that AML technology has kept pace with regulation changes.** | **Count** |
| Disagree | 5 |
| Agree | 4 |
| Neither agree nor disagree | 2 |
| Strongly disagree | 1 |
| **Grand Total** | **12** |

Those who chose "Disagree" self-identified as follows:

| Which of the following best describes your current occupation? | Which of the following best describes your current job level? |
|---|---|
| **Other (please specify)** | Middle Management |
| **Financial crime prevention (Anti-Money Laundering)** | Middle Management |
| **Consultant (Fin Crime)** | Senior Management |
| **Data Scientist or Machine Learning engineer** | Senior Management |
| **Financial crime prevention (Anti-Money Laundering)** | Entry Level |

And those who “Agree”:

| Which of the following best describes your current occupation? | Which of the following best describes your current job level? |
|---|---|
| **Technology vendor (Other)** | Middle Management |
| **Other (please specify)** | Entry Level |
| **MLRO or CCO** | Owner/Executive/C-Level |
| **Consultant (Fin Crime)** | Owner/Executive/C-Level |

There is not a significant grouping between the two choices.

### *Question 17*

Q17. "I would say that new and revolutionary technology is available to those in Anti-Money Laundering".

| Answer Choices | Response Percent | Responses |
|---|---|---|
| Strongly agree | 11.59% | 8 |
| Agree | 49.28% | 34 |
| Neither agree nor disagree | 20.29% | 14 |
| Disagree | 18.84% | 13 |
| Strongly disagree | 0.00% | 0 |
| | Answered | 69 |
| | Skipped | 35 |

The responses to this question, combines well with those who think AI for AML is “meeting expectations” (Q23):

*Table 11 - Q17 + Q23 "Meeting expectations"*

| Do you think that AI for AML is meeting the industry's expectations? | Meeting expectations |
|---|---|
| **I would say that new and revolutionary technology is available to those in Anti-Money Laundering** | **Count** |
| Agree | 18 |
| Strongly agree | 5 |
| Disagree | 2 |
| Neither agree nor disagree | 2 |
| (blank) | 1 |
| **Grand Total** | **28** |

This supports Hypothesis 3: The AML industry is experiencing CD through AI technology innovation.

### *Question 18*

Q18. Have you used AI technologies to help with the regulatory AML burden?

| Answer Choices | Response Percent | Responses |
|---|---|---|
| None of the above | 29.58% | 21 |
| Yes - In configuration | 15.49% | 11 |
| Yes - In reviewing matches | 39.44% | 28 |
| Yes - In "Rule tuning" | 42.25% | 30 |
| Yes - In the analysis of source data | 28.17% | 20 |
| Other (please specify) | 8.45% | 6 |
| | Answered | 71 |
| | Skipped | 33 |

The most common use of AI technologies is currently “Rule tuning”, this is to use AI to configure the AML system’s matching rules to identify customers more accurately and thereby reduce “false positive” matches.

### *Question 19*

Q19. Do you think that these technologies represent good value for money to the business?

| Answer Choices | Response Percent | Responses |
|---|---|---|
| A great deal | 19.72% | 14 |
| A lot | 35.21% | 25 |
| A moderate amount | 29.58% | 21 |
| A little | 14.08% | 10 |
| None at all | 1.41% | 1 |
| | Answered | 71 |
| | Skipped | 33 |

***Question 20***

Q20. Have AI technologies improved your business's ability to fight financial crime?

| Answer Choices | Response Percent | Responses |
|---|---|---|
| I have found them extremely useful | 12.12% | 8 |
| I have found them very useful | 37.88% | 25 |
| They are somewhat useful | 42.42% | 28 |
| AI is not so useful in AML | 7.58% | 5 |
| AI is not at all useful | 0.00% | 0 |
| | Answered | 66 |
| | Skipped | 38 |

Questions 19 and 20 cover the respondent's views on the value benefits of AI technologies. Interestingly, those who select “below expectations” to question 23 still think that AI technologies represent good value for money.

*Table 12 - Q23 "Below expectations" + Q19*

| Do you think that AI for AML is meeting the industry's expectations? | Below expectations |
|---|---|
| **Do you think that these technologies represent good value for money to the business?** | **Count** |
| A lot | 15 |
| A moderate amount | 14 |
| A little | 7 |
| A great deal | 4 |
| None at all | 1 |
| **Grand Total** | **41** |

This supports Hypothesis 1 by demonstrating a positive response to “AI” and “AML” technologies while simultaneously showing concerns about its effectiveness.

### *Question 21*

Q21. Do you think AI can solve AML volume issues (or alleviate them)?

| Answer Choices | Response Percent | Responses |
|---|---|---|
| Extremely helpful | 16.90% | 12 |
| Very helpful | 39.44% | 28 |
| Somewhat confident | 0.00% | 0 |
| Somewhat helpful | 40.85% | 29 |
| Not so helpful | 2.82% | 2 |
| Not at all helpful | 0.00% | 0 |
| | Answered | 71 |
| | Skipped | 33 |

### *Question 22*

Q22. What are your opinions of the current maturity level of AI technologies in AML? (Question uses Technology readiness levels (TRLs))

| Answer Choices | Response Percent | Responses |
|---|---|---|
| 1. Basic principles observed and reported | 16.67% | 11 |
| 2. Technology concept and/or application formulated | 6.06% | 4 |
| 3. Experimental proof of concept | 25.76% | 17 |
| 4. Technology validated in lab conditions | 6.06% | 4 |
| 5. Technology validated in relevant environment | 10.61% | 7 |
| 6. Technology demonstrated in relevant environment | 4.55% | 3 |
| 7. System prototype demonstration in operational environment | 10.61% | 7 |
| 8. System complete and qualified | 1.52% | 1 |
| 9. Actual system proven in operational environment | 13.64% | 9 |
| Other (please specify) | 4.55% | 3 |
| | Answered | 66 |
| | Skipped | 38 |

Technology readiness levels are a maturity model for exploring how close to production technology is. The results here show that most AI technologies for AML are yet to be in full production mode, with most being experimental.

| Other (please specify) |
|---|
| Not au fait enough with AI to comment. |
| It really depends on what you consider AI. Simple programmes are already being used but complex AI is still in development |
| I think it varies widely depending on the type of tech. There's some great tech out there that is already embedded but many opportunities for more to enhance this more holistically. |

### *Question 23*

Q23. Do you think that AI for AML is meeting the industry's expectations?

| Answer Choices | Response Percent | Responses |
|---|---|---|
| Exceeding expectations | 1.43% | 1 |
| Meeting expectations | 40.00% | 28 |
| Below expectations | 58.57% | 41 |
| | Answered | 70 |
| | Skipped | 34 |

Analysis shows that the Q1 grouping who believe "AI for AML" is meeting expectations are mainly technology vendors:

*Table 13 - Q1 "Technology vendor (FinTech)" + Q23*

| Which of the following best describes your current occupation? | Technology vendor (FinTech) |
|---|---|
| **Do you think that AI for AML is meeting the industry's expectations?** | **Count** |
| Meeting expectations | 5 |
| (blank) | 2 |
| Below expectations | 1 |
| **Grand Total** | **8** |

Whereas the MLRO or CCO roles are more split:

*Table 14- Q1 "MLRO or CCO" + Q23*

| Which of the following best describes your current occupation? | MLRO or CCO |
|---|---|

| **Do you think that AI for AML is meeting the industry's expectations?** | **Count** |
|---|---|
| Below expectations | 9 |
| (blank) | 8 |
| Meeting expectations | 7 |
| **Grand Total** | **24** |

### *Question 24*

Q24. Do you think AML technology vendors successfully integrate AI into their products?

| Answer Choices | Response Percent | Responses |
|---|---|---|
| A great deal | 2.86% | 2 |
| A lot | 14.29% | 10 |
| A moderate amount | 44.29% | 31 |
| A little | 34.29% | 24 |
| None at all | 4.29% | 3 |
| | Answered | 70 |
| | Skipped | 34 |

### *Question 25*

Q25. Have you made use of, or heard of, "AI Ethics"

| Answer Choices | Response Percent | Responses |
|---|---|---|
| Extremely familiar | 2.94% | 2 |
| Very familiar | 16.18% | 11 |
| Somewhat familiar | 29.41% | 20 |
| Not so familiar | 26.47% | 18 |
| Not at all familiar | 25.00% | 17 |
| | Answered | 68 |
| | Skipped | 36 |

AI Ethics uses frameworks to consider AI solutions' risks and moral concerns. Considerations such as transparency and explainability. These frameworks provide a balanced and justified approach to AI development. Without AI Ethics, it is possible, even likely, that AI will have risks to individuals and the companies creating them.

### *Question 26*

Q26. Do you think that AI for AML will lead to job losses?

| Answer Choices | Response Percent | Responses |
|---|---|---|
| Very likely | 1.47% | 1 |
| Likely | 30.88% | 21 |
| Neither likely nor unlikely | 23.53% | 16 |
| Unlikely | 38.24% | 26 |
| Very unlikely | 5.88% | 4 |
| | Answered | 68 |
| | Skipped | 36 |

This question gets to the heart of the issue. The responses here are evenly distributed. Insight can be gained into the groupings by considering the roles of those who responded that it was “likely”:

| Which of the following best describes your current occupation? | Which of the following best describes your current job level? |
|---|---|
| **Technology vendor (FinTech)** | Middle Management |
| **Other (please specify)** | Entry Level |
| **MLRO or CCO** | Senior Management |
| **MLRO or CCO** | Senior Management |
| **MLRO or CCO** | Owner/Executive/C-Level |
| **MLRO or CCO** | Middle Management |
| **MLRO or CCO** | Senior Management |
| **MLRO or CCO** | Senior Management |
| **MLRO or CCO** | Senior Management |
| **Financial crime prevention (Anti-Money Laundering)** | Middle Management |
| **Financial crime prevention (Anti-Money Laundering)** | Senior Management |
| **Financial crime prevention (Anti-Money Laundering)** | Intermediate |
| **Financial crime prevention (Anti-Money Laundering)** | Middle Management |
| **Financial crime prevention (Anti-Money Laundering)** | Senior Management |
| **Consultant (Other)** | Senior Management |
| **Consultant (Fin Crime)** | Senior Management |
| **Consultant (Fin Crime)** | Owner/Executive/C-Level |
| **Consultant (Fin Crime)** | Owner/Executive/C-Level |
| **Consultant (Fin Crime)** | Owner/Executive/C-Level |
| **Consultant (Fin Crime)** | Middle Management |
| **Bank employee** | Middle Management |

And those who selected “Unlikely”:

| Which of the following best describes your current occupation? | Which of the following best describes your current job level? |
|---|---|
| **Bank employee** | Senior Management |
| **Consultant (Fin Crime)** | Senior Management |
| **Consultant (Fin Crime)** | Intermediate |
| **Consultant (Fin Crime)** | Owner/Executive/C-Level |
| **Consultant (Other)** | Owner/Executive/C-Level |
| **Consultant (Other)** | Senior Management |
| **Financial crime prevention (Anti-Money Laundering)** | Senior Management |
| **Financial crime prevention (Anti-Money Laundering)** | Middle Management |
| **Financial crime prevention (Anti-Money Laundering)** | Middle Management |
| **Financial crime prevention (Anti-Money Laundering)** | Intermediate |
| **Financial crime prevention (Anti-Money Laundering)** | Middle Management |
| **Financial crime prevention (Anti-Money Laundering)** | Senior Management |
| **Financial crime prevention (Anti-Money Laundering)** | Senior Management |
| **Financial crime prevention (Anti-Money Laundering)** | Middle Management |
| **MLRO or CCO** | Senior Management |
| **MLRO or CCO** | Senior Management |
| **MLRO or CCO** | Senior Management |
| **MLRO or CCO** | Senior Management |
| **Other (please specify)** | Middle Management |
| **Other (please specify)** | Middle Management |
| **Technology vendor (FinTech)** | Intermediate |
| **Technology vendor (FinTech)** | Middle Management |
| **Technology vendor (FinTech)** | Senior Management |
| **Technology vendor (Other)** | Middle Management |
| **Technology vendor (Other)** | Middle Management |
| **Technology vendor (Other)** | Senior Management |

There is a wide distribution of roles in these reply groups, suggesting that experiences of this issue are also widely distributed. This follows Hypothesis 1 that the risks of current-generation AI need to be better appreciated.

### *Question 28*

Q28. What does artificial intelligence ('AI') mean to you?

| | |
|---|---|
| Answered | 19 |
| Skipped | 85 |

| **Responses** |
|---|
| Machine learning |
| The ability to detect unusual patterns in millions of transactions. |
| Machine learning that flushes out the most valuable snippets of information for human assessment |
| Allowing the system to make decisions based on multiple variables |
| A degree of behavioural and adaptive analytics that in this instance looks for patters of both good and potentially suspicious activity |
| Internal application like red flags |
| A very important tool |
| Ai is the use of unsupervised machine learning models to augment existing aml processes |
| Depends on who you ask as people often define it in different ways. Because of this, I define it very broadly to include analytics, machine learning, various automations, etc. |
| A self learning program that develops over time to recognize patterns |
| Machine based intelligence which delivers more accurate, reliable solutions at a much faster rate as compared to humans |
| Machine learning and algorithm development and implementation to better identify instances of potential money laundering and enable deployment of specialist skill sets to key areas |
| Advances technology/algorithms that utilises data in an intelligent way to provide insights/ predictions that cannot be gained through traditional methods and would be typically be done by a human. |
| partly hyped, partly useful (alt least its promises are ...) |
| At a simple level getting computers to make decisions previously taken by humans. Also a range of technology solutions - eg natural language processing, etc. |
| The ability to filter out the noise false positive and enables reviewer to focus on the positive cases |
| It analysis human behaviour and articulate the same in many KYC, Investigation processes etc. |
| A smart and sometimes self-developing way to solve issues in an ICT setting. |
| An automated process that mimics human processing and reduced the redundancy of workflow while focusing on effectiveness. |

A 50-word word cloud of the above answers is as follows:

*Figure 12 - Word cloud of Q28 responses*

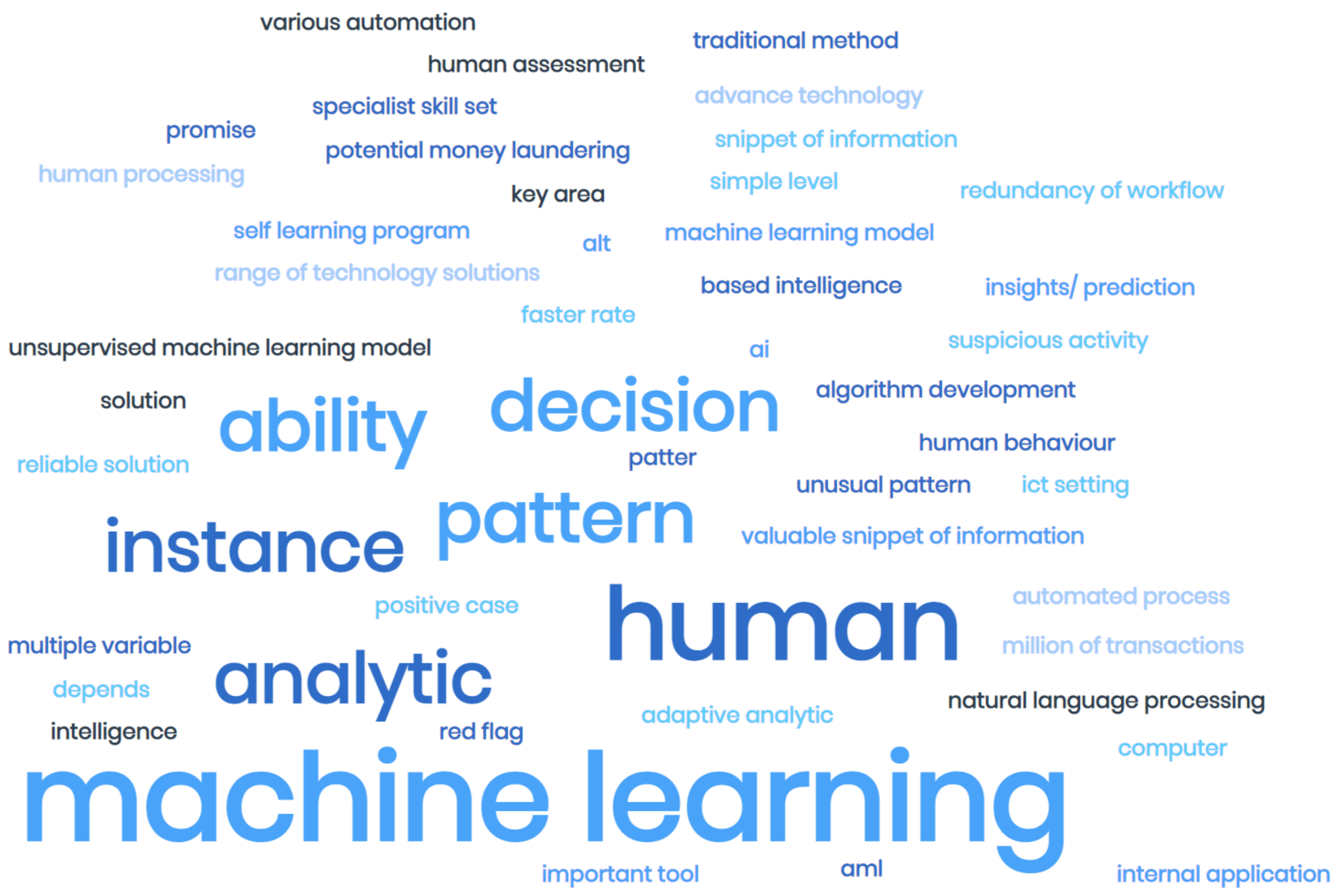

"Machine learning" is the most significant word. This technology is a standard part of the AI umbrella, where machines learn from experience. However, machine learning does not entirely encompass the definition of AI.

### *Question 29*

Q29. How do you perceive AI in the Market?

| | |
|---|---|
| Answered | 18 |
| Skipped | 86 |

| Responses |
| --- |
| Hype - vendors need to cut the hype / advertising a silver bullet. It's useful for some things some of the time, and as long as it's clear where it is used and where not, that's fine. |
| Very beginner. |
| Unsure - Budget constraints in the industry combined with flamboyant sales pitches |
| It's developing |
| Strong in some areas. Weak in others. |
| Yes for the first control line |
| Highly appreciate |
| Buzzword that is rarely implemented properly and is little more than a ruleset |
| Core usage, but needs more players to adapt and adopt |
| It's still in development because KYC is still a relatively young industry. Everything so far has been done manually and discussions about AI, ethics and privacy have been hot. |
| nascent |
| Observed/involved typically for alert predictions and rule tuning. Leveraged effectively can significantly save time and human effort. |
| see above: a lot of hype, but limited proof |
| Developed massively in last few years. Still a big issue translating what it can do against what is needed. The fin crime people don't understand AI and vice versa. |
| Helpful but cannot totally replace the human aspect |
| There are many vendors trying to introduce the AI in the market |
| It's interesting |
| We are just at the beginning and mainly using Machine Learning not yet A.I. |

A 50-word word cloud of the above answers is as follows:

*Figure 13 - Word cloud of Q29 responses*

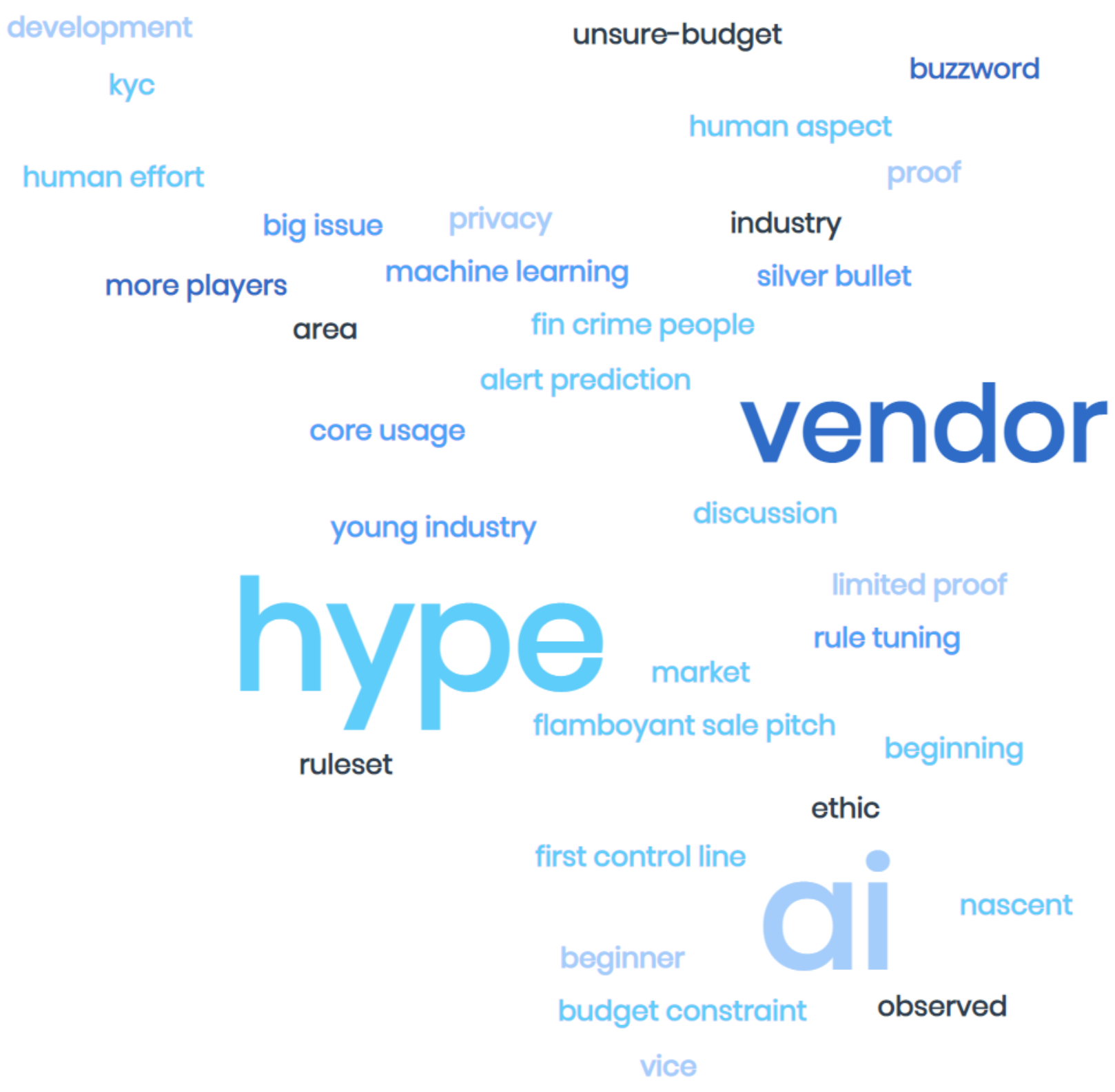

The word cloud shows that the perception of AI has the most significant word of "hype". As discussed in the Factiva results section, hype is the predicted stage for the topic in the information lifecycle. It also suggests that the topic is experiencing high positive bias but that the respondents are aware of this. This follows and supports Hypothesis 1.

**Section D – Extra questions**

## *Question 30*

Q30. Academia suggests a strong correlation between a bank's benefit to the economy and its ability to combat money laundering. Do you agree?

| Answer Choices | Response Percent | Responses |
|---|---|---|
| Strongly agree | 21.74% | 5 |
| Agree | 47.83% | 11 |
| Neither agree nor disagree | 21.74% | 5 |
| Disagree | 8.70% | 2 |
| Strongly disagree | 0.00% | 0 |
| | Answered | 23 |
| | Skipped | 81 |

The results from this question lend public credence to the arguments by Idowu and Obasan (Idowu & Obasan 2012) in the literature review. It also shows the dedication and commitment of those fighting this financial crime.

***Question 31***

Q31. "Crying wolf theory" states that defensive reporting "deluges" investigative units to the point that they cannot effectively act upon them. Do you agree?

| Answer Choices | Response Percent | Responses |
|---|---|---|
| Strongly agree | 26.09% | 6 |
| Agree | 47.83% | 11 |
| Neither agree nor disagree | 21.74% | 5 |
| Disagree | 0.00% | 0 |
| Strongly disagree | 4.35% | 1 |
| | Answered | 23 |
| | Skipped | 81 |

This question’s results support the "Crying Wolf Theory" by Raweh, as detailed in the literature review. It also supports the arguments this paper draws from the Tadesse section of the literature review. It therefore gives strong evidential support to Hypothesis 2, suggesting that defensive reporting acts as a blocker to innovation.

***Question 32***

Q32. How would you rate this questionnaire?

| Answer Choices | 1 | 2 | 3 | 4 | 5 | Total | Weighted Average |
|---|---|---|---|---|---|---|---|
| Stars | 0 | 1 | 2 | 11 | 8 | 22 | 4.18 |
| | | | | | | Answered | 22 |
| | | | | | | Skipped | 82 |

Do you have any comments, questions, or concerns?

| Responses |
|---|
| The average competence in this space needs to rise dramatically. Government can't do it on their own. |
| Would be good to participate in thinktanks on AML advanced solutions. |
| A little subjective in places. Good luck pulling items together. |
| Re: 31. If there was an answer higher than "Strongly agree," I'd pick it. Every bank I've consulted in the past few years (major, top 10 ones, all on Fortune 500) either already was or has been transitioning toward a much more defensive filing route. In some cases, I've seen the number of filed SARs go up by 4-500% in just a year or two for no reason other than being more defensive. |
| Probably the challenges faced in the AML domain apart from Regulatory or AI can be raised as a question |
| My concern is not about AI. My concern is about management and first line employees in the 1st line of defence. They should be more motivated to do the utmost to avoid money laundering and tax evasion. At the basis of that lies a strong ethical choice: what sort of clients to we want to service and which clients do we want to steer clear of. This is apparently a choice that takes courage because I have rarely seen it made. |

This final question asks for feedback on the questionnaire itself. The comment regarding the options in Question 31 is a highlight. While only a single point of evidence, it does suggest a direction further research could develop from this paper.

**Interview findings**

The focus of the interviews was on two main themes:

1. Evidence for hypotheses 1,2, and 3.
2. Questions where the questionnaire results did not clearly signify a single highest supported option. This has been identified as questions: 8, 15, 16, 23 and 26.

The interviews were summarised to protect the identities of the interviewees.

**Interview 1. AML screening professional**

This interviewee works within a team supporting the review of compliance matches generated from name screening. They have seen considerable growth in the region in clients and opportunities thanks to the increase in review matches. It was discussed how “everybody” is impacted by new legislation and its demand upon the regulated. The struggle customers feel is often due to “overlapping” legislations requiring “subjective interpretation”, a situation that customers do not wish to pass on to the end users by asking additional questions when opening the account.

The screening team is often asked directly for guidance by the customers, which is not within their contract to provide. This is thought to be because customers requiring outsourced support are often smaller and not focused on large-scale compliance (like a larger bank). Many customers in such a situation naturally report “permission” SARs to cover themselves from future blame.

They had heard of AI technologies being introduced via smaller start-ups to screen for false positives for customers. They understood the sense of pursuing AI technologies for decision-making, particularly where this would prove better than humans. AI ethics was a recognised but secondary concern, given the need to deal with the volumes. They did not think AI could completely replace humans as it may be “fooled” by overreliance on AIs to solve constantly changing requirements and “subjective”. Many companies would not be able to benefit from a complex AI solution due to a lack of data experts with little compliance and AML knowledge.

**Interview 2. AML technology professional**

This interviewee is an advanced technologist working in a team supporting AML technology implementations. They showed concern for the quality of AI data training sets, which are often passed around publicly and that this could be a vector for hacking AI technologies used in financial crime prevention. They discussed the possibility of internal corruption being missed by fully automated systems compared to humans. They said that AI is seen as a “silver bullet” but is still a “black box” that risks replicating the same problems seen in human screening because humans train them. They mentioned that AI is best seen as “just another tool” in the toolbox. They mentioned a concern with some AI technologies, such as deep learning, which is trained from humans' past performance, and asked, “what if one of them was corrupt?” and deliberately tried to throw off the AI training phase. With such a significant impetus on AI in the AML industry, it was mentioned that the broader concerns about AI technologies being used in other industries, such as BERT in Natural Language Processing, which has demonstrated a potential for misuse. They showed concern for humanity knowing their nature well enough to simulate our thinking processes to the degree required by AI in AML, particularly decision-making. They also showed concern with the privacy implications of AIs and that an AI decision could have large ramifications on human lives without humans in the loop. Finally, they questioned the ability to derive the features from the human activities required to train an AI correctly.

**Interview 3. AML banking professional**

This interviewee is a senior expert in AML compliance. This interviewee noted that they had seen many different AI technologies presented as solutions and was aware of multiple technology teams working on integrating AI into AML. It was discussed how there was a knowledge gap in the AML industry in making the best use of potential AI solutions, leading to a rise in “middlemen” consultants selling services to both sides. This was seen as a symptom of the need of the regulated and evidence for the impact regulation has on the volume of matches and subsequent SARs.

**Interview 4. AML sales professional**

This interviewee is a senior expert in AML and solutions. This person’s primary concern was how to apply AI when talking about compliance AML. They felt AI and machine learning would assist in reducing the burdens of name screening. Most review investigations are to do with the resolution of “linking the dots” between data provided by the regulators (and legal bodies) and the customer data. They noted that machine learning is more valuable than full AI. The need for a balanced approach to AI ethics was discussed and agreed upon. They noted that many solutions using advanced AI, such as neural networks, run into problems with testing and auditing when used within AI. They noted that the target for AI should be where the customer data is “big data” (usually defined as data sets over 5Gb in size) to combat the up to “80%” of false positives leading to SARs. The best technology would focus on the remaining 20% of the data and then simulate human decision-making on this data. However, they showed strong concerns that AIs will make classification mistakes that propagate through the network before they are fixed. This could negatively impact “innocent” customers. They noted that customers tend for the “machine to do everything”, often what is beyond the capabilities of the technology. They also noted the need to develop AI technologies with diverse groups or bias issues would be present.

## Chapter 5. Discussion and conclusion

### Introduction

This chapter will discuss how much the research has addressed the substantive objectives, how far the theoretical objectives have been achieved and the degree to which the findings can be generalised. It will also consider the direct implications of the research findings in the form of policy recommendations.

### Hypotheses validation

All three of the hypotheses tested by the research have had some evidential validation of their formation:

**The risks of current-generation AI need to be better appreciated and understood by those working in the AML industry.** This is clearly shown to be an issue by the hype and sentiment of the news, from the responses to the questionnaire (particularly Q17 combined with Q23), and from the interviewees' responses.

**The regulators and governmental bodies act as "blocking", "protecting", or "safeguarding" devices, stalling innovation.** Yes, somewhat naturally, they do, as shown by the significant impacts on the regulated outlined in the questionnaire, the abundance of SARs, and the discussions with all four interviewees who represent someone from every significant grouping in the questionnaire (Q1).

**The AML industry is experiencing creative destruction through the pursuit of AI.** One of the elements of CD noted in the literature review was that the moral dimension is often ignored. Some interviewees believed that the necessity of solving the burdens in the industry outweighed the ethical dimensions calling for restraint. As described, CD requires capitalism, which is to say competition, to act as the force driving the process. AML is in a quandary, where the instinct to tighten legislation is a huge balance against this force. Two tremendous forces acting against each other, like tectonic plates, are liable to cause massive destruction and fallout when innovation finally wins out and flips the paradigm. The questionnaire shows AI is coming to AML. Whether that has a positive impact or not is down to several policy recommendations.

## Conclusion – Creative Destruction in AML?

Creative destruction was the name Schumpeter gave to a process where innovations, driven by the capitalist "compulsion to accumulate" (Schumpeter, 1976), are created from within the organisation, destined to rise and destroy the dominant paradigm. It can only be seen operating when viewed over time and is constantly kept in check by protecting forces "blocking" change. It happens to almost every industry, and the recent "disruptive start-up" culture is a modern formation. Critical analysis of the existence of this process raises the question of whether something only observable in aggregate is happening and what evidence for that may look like on a more macro scale. This paper has endeavoured to identify methods for obtaining such evidence and researching its presence. From the results, there is an evidential basis for the theory in the following ways:

It appears that AML is experiencing a unique form of CD. As the results show, the blocking forces (the regulators) have forced the industry to seek large numbers of external company employees to keep up with the demands of the regulations. Therefore, it is a natural development to seek technological innovation and reduce the manual burden placed upon the company. The market should be overturned through successful innovations, and the paradigm replaced. And thus, we would see the CD process in action as described.

However, in this case, the blocking forces operate from the highest of motives of protecting society from crime. AML regulation is required to fight money laundering, a crime that enables local criminal activity to have a global economic impact. It attacks financial centres and, as noted by the 2022 review of the UK AML regulatory regime (HMTreasury, 2022), openness to overseas investment brings with it the risk of the UK economy being used for laundering. The regime fighting ML has changed over its 30-year lifetime in response to these threats and weaknesses, attempting to balance crime-fighting without strangling global trade. The creation of FATF also created an approach to AML now seen in nearly all global jurisdictions, but, as this paper has shown – it also led to the same problems arising globally. Therefore, if the UK's answer to these challenges can be said to be ineffective, as reported in the Treasury Committee's Eleventh Report of Session (Mp and Thewliss, 2022), then the global approach to AML is in the same dilemma. As the Committee noted, "There is no "Silver Bullet" solution" (Mp and Thewliss, 2022, p. 14).

A silver bullet is the hope, or perhaps "hype", that AI promises. The Factiva research showed growth in news articles about AI that directly correspond to the hype stages of the information lifecycle, with nary a concern about its applicability. However, AI is not a simple technology to get right, and as the responses to the questionnaire make clear, there needs to be a high level of understanding of what AI is offering. Few AI product integrations have successfully made it out of the conceptual stages. Even FATF recognises that AI in AML only "may" enhance capabilities (FATF, 2021). Their reticence stems from several regulatory "challenges", not least of all the lack of "supervisory acceptance" (FATF, 2021, p. 37). This was highlighted in a post on the FCA regulator website in 2018 (Gruppetta, 2018) where AIs troubled applicability to the volume problems in AML was described as "rational scepticism", which is an interesting comment when the problems themselves are created by the regulations enforced by the very same regulator. An example of CD's innovation blocking-force in action.

Capitalist society needs CD to operate and drive innovations. These innovations drive global economic growth as one party shifts to a new operating paradigm, quickly copied by others looking to remain competitive. In the AML technology industry, the possible economic benefits of the current paradigm have already been realised. It would have already been up-ended if it were not for the protecting activities of the regulatory regimes.

The question is whether the need for innovation, a true societal aim, trumps the potential ethical considerations of using AI to achieve it. If AI enables a real impact on financial crime, does the moral question go away? These are the cutting-edge normative questions AI ethics helps companies evaluate. It recognises that, on the one hand, the lack of AI standards plays into the hands of fully globalised financial criminals. Therefore, as automation increases, the impacts of these crimes could quickly propagate through poorly thought-out AI solutions. However, on the other hand, not innovating in the AML industry threatens to render current methods so ineffective that the global regime faces collapse.

Both paths contain global economic impacts. One path automates a large proportion of jobs in the industry away and potentially creates economic havoc. The other allows innovation through AI, creating a new technological arms race for, as interviewee 2 commented, criminals can use AI too.

Consequently, this paper contends that until a "FATF-like" AI watchdog is created, global risks remain to any industry adopting artificial intelligence at the expense of human activity.

## Limitations of research

This paper recognises that there are limitations in the methods employed. Indeed, a larger audience for the questionnaire would have improved its generalisability. The worldwide MLRO count is undoubtedly above the UK number of approximately 19,000. Thus a survey directed at this more significant number may have resulted in different outcomes or a more regional emphasis. This paper's questionnaire had results from mainly the UK, Europe and the Middle East, with only some respondents in Asia. A more comprehensive view, including China, would have been of interest.

The sole use of Factiva for news data is to ignore its competitors. Moreover, a sentiment score generated by AI could have worked a more extensive data set, however – and given the claims about AI set out in this paper – the use of "off the shelf" sentiment algorithms are known to the author to have inferior and unreliable results with news data. Unfortunately, there was no time in the project to produce a working engine.

In interviewing people directly, this paper does not contend that their thoughts are uniquely generalisable, following the rule that the "plural of anecdote is not data" (Mieder, Shapiro and Doyle, 2012). However, their opinions and observations are correlated with those found in the questionnaire, as are the attitudes in the literature review. A larger group of interviews would have been of benefit, should time have been allotted.

Future researchers may find the above limitations helpful when designing their research methods. It may be that AI in AML is nascent, and the impacts are not yet evident regarding job losses. The author was unable to find significant mentions of such losses within Factiva but is personally aware of many thousands. Speculation is that the news currently is too optimistic about AI in AML to record them. The author attempted to interview several editors regarding their articles, but they all refused to be interviewed on the record.

## Policy Recommendations

When considering AI as a potential solution to AML, the research undertaken by this paper forwards several primary considerations:

1. There is an ethical dimension to all innovations, and in AI, this is AI Ethics. To consider AI Ethics is not to block innovation at all. It is instead to use a logical moral framework to prevent fallout from the global lack of standards. As a result, ethical AI is safe, with less risk to the company and the industry.
2. As shown in the literature review, the transparent reporting of SARs identifies threats to the industry. However, the over-reporting of SARs to avoid the potential downsides actively harms the industry. Some AI solutions will assist with this. As interviewee 2 puts it, “tools in the toolbox”. However, AI is not a silver bullet, and, as interviewee 3 notes, “middlemen” consultants are a symptom that the complexity of AI equals the complexity of AML regulation.

3. While technology, and the CD process, target expensive manual burdens in any business paradigm, it does not address the AML industry's fundamental issues and its relationship with regulated companies and regulators. AI may be more a salve than an answer to AML screening, but true innovation needs to come at the regulator level. As a result, the requirements of the SAR need to be reimagined. Alternatively, as this paper's research shows, the burden on the regulated will deleteriously impact their ability to answer that regulation satisfactorily. And that may lead to another type of destruction altogether.

## List of Figures